\documentclass[12pt,preprint]{aastex}
\newcommand{\hii}{\mbox{\ion{H}{2}}}
\newcommand{\ha}{\mbox{H$\alpha$}}
\newcommand{\hb}{\mbox{H$\beta$}}

\newcommand{\heii}{\mbox{\ion{He}{2}~$\lambda$4686}}
\newcommand{\oiiir}{\mbox{[\ion{O}{3}]~$\lambda$5007}}

\newcommand{\oii}{\mbox{[\ion{O}{2}]~$\lambda$3726+3729}}
\newcommand{\niil}{\mbox{[\ion{N}{2}]~$\lambda$6548}}
\newcommand{\niir}{\mbox{[\ion{N}{2}]~$\lambda$6584}}
\newcommand{\nii}{\mbox{[\ion{N}{2}]~$\lambda$6548+6584}}
\newcommand{\sii}{\mbox{[\ion{S}{2}]~$\lambda$6717+6731}}
\newcommand{\siil}{\mbox{[\ion{S}{2}]~$\lambda$6717}}
\newcommand{\siir}{\mbox{[\ion{S}{2}]~$\lambda$6731}}
\newcommand{\siii}{\mbox{[\ion{S}{3}]~$\lambda$9531}}

\newcommand{\m}{\mbox{$^{-1}$}}
\newcommand{\mm}{\mbox{$^{-2}$}}
\newcommand{\mmm}{\mbox{$^{-3}$}}
\newcommand{\sci}[3]{\mbox{$#1. #2\cdot  10^{#3}$}}
\newcommand{\scipm}[3]{\mbox{$(#1\pm #2)\cdot  10^{#3}$}}
\newcommand{\taurad}{\mbox{$\tau_{\rm rad}$}}
\newcommand{\taubal}{\mbox{$\tau_{\rm Bal}$}}
\newcommand{\taubalp}{\mbox{$\tau^{\prime}_{\rm Bal}$}}
\newcommand{\taucov}{\mbox{$\tau_{\rm cov}$}}
\newcommand{\dens}{\mbox{$n_{\rm e}$}}
\singlespace
\slugcomment{To appear in the September 2004 issue of the Astronomical Journal}

\begin{document}

\title{NGC 604, the Scaled OB Association (SOBA) Prototype. I: Spatial 
       Distribution of the Different Gas Phases and Attenuation by Dust}
\shorttitle{The Gas Phases of NGC 604}

\author{J. Ma\'{\i}z-Apell\'aniz\altaffilmark{1,2},
        E. P\'erez\altaffilmark{3}, \&
        J. M. Mas-Hesse\altaffilmark{4}}

\altaffiltext{1}{Space Telescope Science Institute, 3700 San Martin Drive,
Baltimore, MD 21218, U.S.A.}
\altaffiltext{2}{Affiliated with the Space Telescope Division of the European 
Space Agency, ESTEC, Noordwijk, Netherlands.}
\altaffiltext{2}{Instituto de Astrof\'{\i}sica de Andaluc\'{\i}a (CSIC),
P. O. Box 3004, 18080 Granada, Spain.}
\altaffiltext{3}{Laboratorio de Astrof\'{\i}sica Espacial y F\'{\i}sica
Fundamental - INTA, P. O. Box 50727, E-28080 Madrid, Spain.}

\begin{abstract}

We have analyzed HST and ground-based data to characterize the different
gas phases and their interaction with the Massive Young Cluster in NGC 604,
a Giant \hii\ Region in M33. The warm ionized gas is made out of two
components: a high-excitation, high-surface brightness \hii\ surface
located at the faces of the molecular clouds directly exposed to the
ionizing radiation of the central Scaled OB Association; and a
low-excitation, low-surface brightness halo that extends to much larger
distances from the ionizing stars. The cavities created by the winds and SN
explosions are filled with X-ray-emitting coronal gas. The nebular lines
emitted by the warm gas experience a variable attenuation as a consequence
of the dust distribution, which is patchy in the plane of the sky and with
clouds interspersed among emission-line sources in the same line of
sight. The optical depth at \ha\ as measured from the ratio of the thermal
radio continuum to \ha\ shows a very good correlation with the total CO
($1\rightarrow 0$) column, indicating that most of the dust resides in the
cold molecular phase. The optical depth at \ha\ as measured from the ratio
of \ha\ to \hb\ also correlates with the CO emission but not as strongly as
in the previous case.  We analyze the difference between those two
measurements and we find that $\lesssim 11\%$ of the \hii\ gas is hidden
behind large-optical-depth molecular clouds; we pinpoint the positions in
NGC 604 where that hidden gas is located. We detect two candidate compact
\hii\ regions embedded inside the molecular cloud; both are within short
distance of WR/Of stars and one of them is located within 16 pc of a
RSG. We estimate the age of the main stellar generation in NGC 604 to be
$\approx$~3~Myr from the ionization structure of the \hii\ region, a value
consistent with previous age measurements. The size of the main cavity is
smaller than the one predicted by extrapolating from single-star wind-blown
bubbles; possible explanations for this effect are presented.
\end{abstract}

\keywords{dust, extinction --- HII regions ---
ISM: individual (NGC 604) --- ISM: structure --- 
Galaxies: ISM}

\section{INTRODUCTION}

        Giant \hii\ Regions (GHRs) constitute one of the most conspicuous
type of objects in spiral and irregular galaxies at optical
wavelengths. For example, the 30 Doradus nebula can be detected with the
naked eye at a distance of 50 kpc and five GHRs in M101 have their own NGC
number at a distance of 7.2 Mpc. GHRs are powered by Massive Young Clusters
(MYCs) that include $10^2-10^3$ O and WR stars, as opposed to the few stars
of this type encountered in regular \hii\ regions\footnote{For example, the
Trapezium cluster in the Orion nebula includes only one O star, $\theta^1$
Ori C; the rest of the massive stars in the cluster are actually early-type
B stars that provide only a small contribution to the total number of
ionizing photons \citep{ODel01}.}.  The advent of high-spatial resolution
data has revealed that GHRs have complex structures and cannot be
accurately modeled as simple Str\"omgren spheres centered on the ionizing
cluster. The deposition of energy in the form of ultraviolet light, stellar
winds, and supernova explosions and the complex initial distribution in
mass and velocity of the ISM produce a spatial distribution of the warm
ionized gas that is anything but spherically symmetric. The analysis of the
best studied GHR, 30 Doradus, reveals that most of the optical nebular
emission originates at the surface of a molecular cloud in those points
where it is directly exposed to the UV flux from the central cluster. The
molecular cloud is what remains from the original material that gave birth
to the MYC and the resulting region of intense optical nebular emission is
a relatively thin layer instead of a classical Str\"omgren sphere
\citep{Walbetal02a,Rubietal98b,Scowetal98}.  This near-bidimensional
character of an \hii\ region is also observed in other objects, from the
low-mass Orion nebula (\citealt{Ferl01} and references therein) to much
more massive regions such as N11 in the LMC \citep{Barbetal03}. Another
property shared by these objects is that a second thin layer, the Photo
Dissociation Region (PDR), where the non-ionizing UV light transforms H$_2$
into atomic hydrogen, can be found sandwiched between the \hii\ region and
the molecular cloud \citep{HollTiel97}.

        The MYCs at the center of GHRs can be divided into two types: Super
Star Clusters (SSCs) are organized around a compact ($1-3$ pc) core while
Scaled OB Associations (SOBAs) lack such a structure and are more extended
objects, with sizes larger than 10 pc \citep{Maiz01b}. The cores of some
SSCs (e.g. 30 Doradus) are surrounded by extended halos that are themselves
similar to SOBAs in terms of structure and number of stars.  This
distinction between SSCs and SOBAs is important for the long-term evolution
of the cluster, since the extended character of the latter makes them
highly vulnerable to disruption by tidal forces on time scales shorter than
the age of the Universe. Therefore, only SSCs are likely to survive to
become the globular clusters of the future \citep{FallRees77}. MYCs are
expected to form in a time scale shorter than 10 Myr but some of the
best-studied examples, such as 30 Doradus and N11, reveal that their stars
are not completely coeval: several million years after the central regions
have been formed, there is induced star-formation in their periphery,
leading to the so-called two-stage starbursts
\citep{Parketal92,WalbPark92,Walbetal99}.

The analysis of GHRs in galaxies experiencing strong episodes of star
formation has been the basic tool for decades to understand the properties
of starbursts and their interaction with the surrounding medium.  In order
to characterize GHRs correctly, one has to establish the balance between
the ionizing UV flux produced by the stars and the flux of the Balmer lines
emitted by the GHR. The former can only be obtained from the analysis of
the stellar continua in the UV domain, where the extinction is strongest.
The latter fluxes are also affected by extinction, and it has been known
for a long time that a simple correction using the \ha/\hb\ flux ratio
might underestimate the intrinsic fluxes, due to the complex geometry of the
dust and the gas within the GHR \citep{CaplDeha86}. Furthermore, attempts
to study the extinction in starbursts using integrated data have discovered
that, on average, the color excess $E(B-V)$ experienced by the stars is
about half that experienced by the gas \citep{Calzetal00}, therefore
complicating even more the calculation of the balance between the ionizing
photons and the observed nebular spectrum, and making it very difficult to
derive the fraction of ionizing photons escaping from the GHR. 

The main objective of this work is therefore to characterize the
properties, distribution and interrelation between the stars, gas and dust
components of GHRs, in order to understand how they affect the resulting
spectral energy distributions at various wavelength ranges. To achieve this
goal we started some years ago a program to analyze in detail nearby
resolved GHRs. In this work we present the first results obtained on NGC
604, concerning the properties of its gas and dust phases.

        NGC 604 is a Giant \hii\ Region in M33, the third largest galaxy in
the Local Group after M31 and the Milky Way, located at a distance of 840
kpc \citep{Freeetal01} along a direction with low foreground extinction
\citep{GonDPere00}. 30 Doradus and NGC 604 are the two largest GHRs in the
Local Group. NGC 604 is powered by a MYC without a central core that
contains $\gtrsim 200$ O and WR stars \citep{Maiz01b,Huntetal96,
Drisetal93}. The MYC is located inside a cavity that has the kinematic
profile of an incomplete expanding bubble.
On the other hand, the brightest knots all have kinematic profiles that can
be well characterized by single gaussians \citep{Sabaetal95,Maiz00}, a
strong indication that the overall dynamics of the GHR is not yet dominated
by the kinetic energy deposition by stellar winds and SN explosions
\citep{TenTetal96}.

Given its proximity, size, low foreground extinction and
structural/kinematical properties, NGC 604 constitutes an ideal object for
our study. It can indeed be considered as the prototype Scaled OB
Association. We have combined HST and ground-based data to characterize the
Massive Young Cluster, the surrounding Interstellar Medium and the
interaction between them. In this first article we present and discuss the
properties of the GHR created by the SOBA and its relationship with the
molecular and coronal gas and dust clouds in the region. We have used
WFPC2/HST and ground-based data to characterize the spatial distribution of
the optical nebular emission properties. We have complemented this
information with published data at other wavelength ranges. In a subsequent
paper we will analyze the properties of the individual massive stars in NGC
604 and its extinction using HST imaging and HST objective-prism
spectroscopy, aiming to derive accurately the properties of the individual
stars and their evolutionary state. This will allow to compare the ionizing
flux being emitted in the region with the value derived in this work from
the nebular emission, constraining so the fraction of ionizing photons
potentially escaping the region without contributing to the ionization.

        We present the observations and describe the data analysis in
section 2.  We describe the attenuation by dust experienced by the warm
ionized gas in section 3 and its density and excitation structure in
section 4. Finally, we discuss our results in section 5.

\section{OBSERVATIONS AND DATA ANALYSIS} 

\subsection{WHT Long-slit spectra}

\begin{figure}
\centerline{\includegraphics*[angle=90,width=\linewidth]{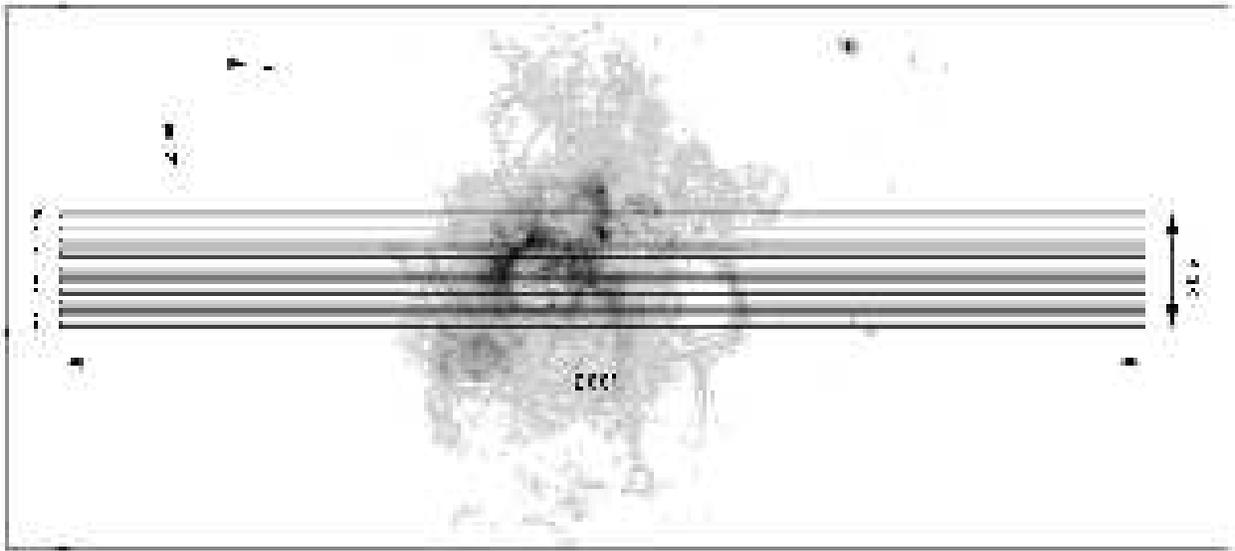}}
\caption{The ten long-slits superimposed on one of the F656N WFPC2 images. Note
that north is towards the bottom and that this orientation is used throughout
the article.}
\label{longslits}
\end{figure}

        We obtained optical long-slit spectra of NGC 604 on the 18-19 August 
1992 night with the ISIS spectrograph of the William Herschel Telescope 
as a part of the GEFE collaboration\footnote{GEFE (Grupo de Estudios de 
Formaci\'on Estelar) is an international group whose main objective is the 
understanding of the parameters that control massive star formation in 
starbursts. 
GEFE obtained 5\% of the total observing time of the telescopes of the 
Observatories of the Instituto de Astrof\'{\i}sica de Canarias. This time is 
distributed by the Comit\'e Cient\'{\i}fico Internacional among international 
programs.}. The spectra were obtained at ten different positions, all of them 
at a P.A. of 90\arcdeg, with an effective width for each one of 1\arcsec\ and a 
spacing of $2\arcsec-3\arcsec$  between the center of each two 
consecutive positions (Fig.~\ref{longslits}). Two spectra were taken 
simultaneously at each position, one in the range between 6390 \AA\ and 6840 
\AA\ (red arm) and the other one in the range between 4665 \AA\ and 5065 \AA\ 
(blue arm), both of them with a dispersion of approximately 0.4 \AA\ / pixel. 
The slit was approximately 200\arcsec\ long, with a spatial sampling along the 
slit of 0\farcs 33525 / pixel in the red arm and 0\farcs 3576 / pixel in the 
blue arm. The exposure times ranged between 900 and 1200 seconds and the 
airmass varied between 1.21 and 1.01. More details of the data reduction and 
analysis are given in \citet{Maizetal98} and \citet{Maiz00}. 

     With this strategy the direction perpendicular to the slit was 
under-sampled. However, taking into account that the seeing was comparable to the
$1\arcsec -2\arcsec$ not sampled between subsequent slits, linear interpolation 
over the areas not covered is enough to reproduce the spatial distribution of 
the different parameters. This approach allowed us to cover the whole central 
region of NGC 604 in less than a single night without losing any significant 
amount of 
information. The reconstructed \ha\ image (top panel in Fig.~\ref{slosrmaps})
shows indeed all the features obtained with imaging techniques at similar 
resolution. These data have been used for a variety of purposes in several
previous works \citep{TerEetal96,Maiz00,TenTetal00}. Here we will use them 
to produce 2-D maps of the \siil /\siir\ ratio across the face of NGC 604 and
to calibrate the nebular WFPC2 data.

An additional set of spectroscopic data with the WHT were obtained 
by means of a spatial scan of the spectrograph 
perpendicular to the slit length direction, which was positioned
at a PA=60\arcdeg. These scanned spectra were centered at 
RA = 1$^{\rm h}$ 31$^{\rm m}$ 43$^{\rm s}$, 
dec = 30\arcdeg\ 31\arcmin 52\arcsec, and they cover the core 
of the region by displacing a 1\arcsec-wide long-slit in steps of 1\arcsec\ 
and taking at each position a 1 minute exposure. The details of this 
observation and of its calibration are given in \citet{GonDPere00}.

\subsection{WFPC2 imaging}

        We list in Table~\ref{wfpc2data} all the NGC 604 WFPC2 data currently 
available in the HST archive. The data correspond to four different programs,
each with a different field of view, but all images include the region where
the nebular line emission caused by the NGC 604 central cluster can be clearly
identified. The images available in each program can be summarized as follows: 
5237, optical broad-band ($UVI$) + \ha; 5384, FUV broad-band; 5773, 
``standard'' narrow-band (\oiiir, \ha, and \sii) + wide Str\"omgren $y$; and 
9134, ``extended'' narrow-band (\oii, \hb, \niir, and \siii). Only some of those
images will be used in this paper but the common part of the reduction process 
is described here in order to use the data in a subsequent paper.

\begin{deluxetable}{llcccc}
\tablecaption{NGC 604 WFPC2 data available in the HST archive.\label{wfpc2data}}
\tabletypesize{\small}
\tablewidth{0pt}
\tablehead{\colhead{Prog.} & \colhead{P.I.} & \colhead{Filter} & \colhead{Band} & Data sets & Exp. times (s)}
\startdata
5237 & Westphal & F336W & WFPC2 $U$             & u2ab0207t + 208t          &    600 +    600                                  \\
     &          & F555W & WFPC2 $V$             & u2ab0201t + 202t          &    200 +    200                                  \\
     &          & F814W & WFPC2 $I$             & u2ab0203t + 204t          &    200 +    200                                  \\
     &          & F656N & \ha                   & u2ab0205t + 206t          & 1\,000 + 1\,000                                  \\
5384 & Waller   & F170W & Far UV                & u2c60b01t + 202t          &    350 +    350                                  \\
5773 & Hester   & F502N & \oiiir                & u2lx0305t + 306t          & 1\,100 + 1\,100                                  \\
     &          & F656N & \ha                   & u2lx0301t + 302t          & 1\,100 + 1\,100                                  \\
     &          & F673N & \sii                  & u2lx0303t + 304t          & 1\,100 + 1\,100                                  \\
     &          & F547M & Wide Str\"omgren $y$  & u2lx0307t + 308t          &    500 +    500                                  \\
9134 & Garnett  & F375N & \oii    & \hspace{2.5mm}u6cj0201m + 202r +        & \hspace{4mm}2\,700 + 2\,700 +                    \\
     &          &       &              & \hspace{8.5mm}203m + 301m          & \hspace{4mm}2\,700 + 2\,700\hspace{4mm}          \\
     &          & F487N & \hb                   & u6cj0101m + 102m + 103m   & 2\,700 + 2\,700 + 2\,700                         \\
     &          & F658N & \niir                 & u6cj0104m + 105m          & 1\,300 + 1\,300                                  \\
     &          & F953N & \siii   & \hspace{4.0mm}u6cj0401m + 402m + 403m + & \hspace{4mm}1\,300 + 1\,300 + 1\,300 +           \\
     &          &       &              & \hspace{8.5mm}404m + 405m + 406m   & \hspace{4mm}1\,300 + 1\,300 + 1\,300\hspace{4mm} \\
\enddata
\end{deluxetable}

        The standard WFPC2 pipeline process takes care of the basic data
reduction (bias, dark, flat field corrections). Cosmic rays were eliminated
in each case using the {\tt crrej} task and the correction of the geometrical
distortion and montage of the four WFPC2 fields was processed using the 
{\tt wmosaic} task, both of which are included in the standard STSDAS package 
for WFPC2 data analysis. Hot pixels were eliminated using a custom-made IDL
routine. 

The next step in the data reduction, the registering of the
images obtained under different programs, followed a more elaborated 
process. The geometric distortion of the WFPC2 detectors is well known
\citep{CaseWigg01}, allowing for a precise relative astrometry. However, 
the absolute
astrometry has two problems: First, the Guide Star Catalog, which is used as a
reference, has typical errors of $\sim 1\arcsec$ \citep{Russetal90}. Second, if
only one point in one of the four WFPC2 chips can be established as a precise 
reference using an external catalog, the average error in the orientation 
induces an error of $\sim 0\farcs03$ in a 
typical position in the other three chips.
In our case, a star that is present in all 13 WFPC2 fields 
is included in the Tycho-2 catalog as entry 2\,293-642-1. We used the VizieR 
service \citep{Ochsetal00} to obtain its coordinates and proper motion using the
J2000 epoch and equinox and we then corrected for the different epoch of the
observations. The star is saturated (in some filters only weakly, with only
a few pixels affected, but in others quite heavily, with significant bleeding 
to pixels in the same column) in all filters
except F170W but we were able to fix its centroid within one quarter of a 
WF pixel in all cases using the diffraction spikes. Thus, the procedure followed
to establish a uniform coordinate system had three steps. First, we corrected
for the difference in plate scale for each filter with respect to F555W 
using the parameters provided by \citet{Dolp00a}. Note that at the 
present time the STSDAS tasks {\tt wmosaic} and {\tt metric} use the 
non-wavelength-dependent Holtzmann solution, which can introduce 
$\sim 0\farcs1$ errors for FUV data; errors are an order of magnitude or
less smaller if only optical data are involved.
Second, we rotated all mosaiced images in order to make them share a common 
orientation. Third, using TYC 2\,293-642-1 as a reference, we corrected for the 
general displacement in RA and declination and compared the positions of the 
stars in the central region of NGC 604 (at a distance of $\sim 70\arcmin$) 
using images obtained under 
different filters. As expected, coordinates differed by a few hundredths of an
arcsecond, which we take to be the precision of our absolute 
astrometry\footnote{Recently, \citet{AndeKing03a} and \citet{Kozhetal03} have 
attacked the geometric distortion solution of the WFPC2 with success, improving 
its accuracy significantly, but we do not use those results here since the 
precision we attain is sufficient for our needs.}. 

        In this first paper we are interested in obtaining ``pure nebular''
\ha, \hb, \oiiir, \niir\, and \sii\ images from the narrow-band F656N,
F487N, F502N, F658N, and F673N WFPC2 data, respectively. In order to do
that, we used the broad-band images to eliminate the continuum
contribution.  Following this approach one has to be careful since
broad-band images can be contaminated by the nebular emission itself (this
is especially so for F555W data). If one has to use a heavily-contaminated
broad-band filter to subtract the continuum, the full linear system of
equations for the reciprocal contributions has to be solved (see,
e.g. \citealt{MacKetal00}). Here we used F814W, F547M, and F336W images,
which have much weaker nebular contaminations and can be considered to a
first order approximation as free from contributions from emission
lines. The continuum at \ha, \niir, and \sii\ was interpolated from the
F547M and F814W data while that of \hb\ and \oiiir\ was interpolated from
the F336W and F547M data. In principle, the continuum subtraction for the
first three lines is expected to be more accurate than for the last two for
two reasons: (1) The difference in flux between F547M and F814W is
relatively small for the type of sources that contribute to the continuum
(early-type stars for the central cluster region, nebular
continuum\footnote{Note that, in any case, the nebular contribution is
rather unimportant and in most cases of the same or smaller magnitude as
the uncertainties derived from the photometric calibration.} for the
nebulosity, late-type stars for the background) in comparison to the
difference in flux between F336W and F547M for some of those types.  (2)
Some of those source types have strong Balmer jumps in their spectra, which
makes interpolation between F336W and F547M more inaccurate. With those
caveats in mind, we tested several interpolation mechanisms and the best
results (for both point sources and background) were obtained using a
double (in $\lambda$ and in flux) logarithmic scale, which is exact if the
flux follows a power law in $\lambda$.  We should point out that, even
though the remaining flux due to the continuum contribution at the location
of a bright star in the final nebular image typically averages to zero
(i.e. the overall continuum subtraction is correct), there is usually some
structure visible at such locations in the resulting image. The main reason
for this effect is the undersampled nature of the WFPC2 PSF, which does not
permit a perfect subtraction of one image from another\footnote{There are,
however, at least two more effects, detector saturation and emission
associated with the star itself, which are also important in a number of
cases.}. For that reason, we decided to blank out in the nebular images
those areas with bright stars and when measuring the integrated nebular
fluxes we have interpolated in those regions using the neighboring
pixels. Another issue related to the obtention of pure nebular images, the
mutual contamination between \ha\ and \nii\ in the F656N and F658N filters
is discussed later.

        Another topic that needs to be addressed here is that of the absolute
photometric calibration of the WFPC2 nebular filters. It is stated in the HST
Data Handbook \citep{wfpc2data} that their accuracy is estimated to be 
$\lesssim 5\%$. A recent study by \citet{Rubietal02} found this to be true for 
three of the filters used here, F487N, F502N, and F656N (see
Table~\ref{wfpc2cal}). We performed an independent calibration of four of the
nebular filters (F487N, F656N, F658N, and F673N) using the NGC 604
\ha\ and \hb\ images published by \citet{Boscetal02}, which were kindly made 
available to us by Guillermo Bosch, and our WHT long-slit spectra. For each
filter we used {\tt synphot} \citep{synphot} to obtain the conversion factors 
between detector and physical units both for a continuum source and for an
emission line with the appropriate blueshift corresponding to NGC 604.

\begin{enumerate}
  \item For F487N, we compared the continuum-subtracted integrated flux over
        a $50\arcsec\times 50\arcsec$ region (after smoothing the image with a
        Moffat kernel to degrade its resolution) with the same value derived 
        from the \citet{Boscetal02} data. 
  \item For F656N, we used a similar but more complex procedure. The photons
        detected in that filter can originate in the continuum, \ha\ itself, 
        or one of the neighbor [N\,{\sc ii}] lines. Given that the expected
        ratios are rather low (\niir/\ha $\sim$ 0.1, \niil/\ha $\sim$ 0.03) and
        that the throughput peaks near the wavelength of \ha, the
        contamination by [N\,{\sc ii}] is expected to be not too important
        ($< 10\% $) but still significant. On the other hand, F658N is also
        contaminated by \ha\ emission to a similar degree (\ha\ falls at the
        tail of the filter throughput at the blueshift of NGC 604). Therefore,
        we decided to eliminate the mutual contamination by establishing the
        2 linear equations with the coefficients determined from {\tt synphot}
        and solving for the two unknowns, \ha\ and \niir, using 
        \niir/\niil\ = 3. We then compared the obtained \ha\ flux with the
        \citet{Boscetal02} data in a manner analogous to what we did previously
        for \hb.
  \item To verify the calibration for F658N and F673N we first obtained the 
        continuum-subtracted images (and for the case of F658N we also
        eliminated the \ha\ contribution as detailed in the previous point) and
        degraded their resolution with a Moffat kernel. We then calibrated the 
        \ha\ fluxes from the long-slit spectra with the \ha\ image from 
        \citet{Boscetal02} after shifting the relative positions of the
        long-slits to match the structures seen in the corresponding cuts of
        the images. We used the absolute calibration thus provided to obtain
        the corresponding \niir\ and \sii\ fluxes for the long-slit spectra and 
        then compared the integrated fluxes with those obtained from the WFPC2
        F658N and F673N data.
\end{enumerate}
        
        Given the mutual influence in the calibration of F656N and F658N, the
above procedure was iterated several times until convergence was reached. The
results are shown in Table~\ref{wfpc2cal} in the form of the ratio
between the values of the EMFLX parameter (the factor used to convert from 
ADU/s to monochromatic flux) as provided by {\tt synphot} and as measured here. 
For all four cases we detect that {\tt synphot} slightly overestimates the
throughput of the nebular filters, which is the same effect that was detected
by \citet{Rubietal02} for two of them, F487N and F656N (they did not analyze
F658N and F673N). Our values are similar but not identical, which is not
unexpected given that the different radial velocities of the objects studied is
sufficient to shift the wavelengths of the emission lines by several Angstroms.
In the rest of this paper we will use the calibration discussed here for 
F487N, F656N, F658N, and F673N and the one obtained by \citet{Rubietal02} for
F502N.

        It should be noted that Charge Transfer Efficiency (CTE) effects are 
not included in the results
of Table~\ref{wfpc2cal}, either for the new values or for those extracted from
\citet{Rubietal02}, and that could be a major factor in the existing
differences\footnote{A sign pointing in that direction is that the largest
correction is found for F658N; those exposures have lower backgrounds and were 
obtained later than F656N or F673N, both of which have smaller corrections.
This is what would be expected if CTE effects were playing a significant role.}
Since our data have been directly calibrated using ground-based fluxes (except
for F502N) this should not be a problem for the results in this paper. However,
one should be careful when using the results in Table~\ref{wfpc2cal} with other
data and, if possible, should attempt a similar procedure as the one outlined
in the previous paragraphs\footnote{In a recent paper, \citet{Calzetal04} did a
similar analysis for four different targets. They found object-to-object
variations in the calibration, likely caused by differences in redshift, among
other effects. See also the work by \citet{ODelDoi99} for a calibration of
the nebular filters applicable to resolved Galactic \hii\ regions.}.

\begin{deluxetable}{lccccc}
\tablecaption{Absolute flux calibration of the nebular WFPC2 filters. The
quantity plotted in each case is the EMFLX
parameter (conversion factor between count rate and monochromatic line flux) 
provided by {\tt synphot} \citep{synphot} divided by the value of EMFLX 
measured by alternative calibrations. A value greater than one indicates that
{\tt synphot} overstimates the throughput and, therefore, that the sense of the
correction is to increase the line flux (since in that case a larger real
count rate is necessary to produce a given monochromatic 
flux).\label{wfpc2cal}}
\tabletypesize{\small}
\tablewidth{0pt}
\tablehead{\colhead{Data} & \colhead{F487N} & \colhead{F502N} & \colhead{F656N}
& \colhead{F658N} & \colhead{F673N}}
\startdata
{\tt synphot} calibration       & 1.000 & 1.000 & 1.000 & 1.000 & 1.000 \\
Table 1 of \citet{Rubietal02}   & 1.032 & 0.959 & 1.037 & ----- & ----- \\
\citet{Boscetal02} data         & 1.012 & ----- & 1.101 & ----- & ----- \\
\citet{Boscetal02} + long-slits & ----- & ----- & ----- & 1.210 & 1.114 \\
\enddata
\end{deluxetable}

\subsection{Other data}

        In order to provide a more complete picture of the ISM in NGC 604 we
obtained some additional data from the literature and other archives. Ed
Churchwell generously provided us with the data shown in Fig.~2 of
\citet{ChurGoss99}. Those authors obtained a 3.6 cm
radio continuum map of NGC 604 with
the VLA and combined the data with a ground-based \ha\ image of the region to
produce a map of \taurad, the (true) optical depth experienced by \ha\ 
emission (see Appendix). Here we will use the \taurad\ map to compare it with a
\taubal\ map, that is, a map of the optical depth at \ha\ as measured from the
ratio of the fluxes of the \ha\ and \hb\ emission lines. A vital step to do
an study of the attenuation by dust 
is to obtain a correct registration between the images
obtained at different wavelengths. In order to do this we first degraded the
spatial resolution of our WFPC2 \ha\ and \hb\ images in order to match that
of the radio data. As it is discussed later in the paper, some 
of the bright \hii\ knots are located on extinction gradients and this can
cause a significant displacement between the positions of the intensity peaks
in \ha\ and radio, rendering those knots useless for registration purposes.
Therefore, we decided to use instead for registration (a) the SNR present in 
NGC 604 (knot E in \citealt{ChurGoss99}), which is located in a relatively
dust-free area and should have the same coordinates in both the optical and
radio data \citep{DOdoetal80,Gordetal98}; 
and (b) the intensity minima determined by the cavities in the ionized gas, 
since those structures are also located on areas with low attenuation. This 
process yielded a registration correct to better than 0\farcs 5
without requiring the use of a rotation between the optical and radio data, as
estimated from the residuals in the fit. Given that the radio beam had a HPBW
of $4\farcs 18\times 3\farcs 63$, the precision in the registering is good
enough for our purposes.

        We also extracted from the M33 CO ($1\rightarrow0$) 
survey of \citet{Engaetal03} 
a map of the NGC 604 region. That survey was obtained with BIMA and has a
spatial resolution of 13\arcsec. Finally, we also retrieved from the 
Chandra archive a 90 ks image of NGC 604 obtained with ACIS (P.I.: Damiani). 
The X-ray data were registered to the
optical data using the SNR in NGC 604\footnote{The registering of
the CO and X-ray data is not as vital as that of the radio continuum, since in
those cases we do not calculate intensity ratios using data from the different
wavelength ranges.}.
Here we will use the CO and X-ray data to compare the morphologies of the 
different phases of the ISM in NGC 604. 

\section{NEBULAR MORPHOLOGY AND ATTENUATION BY DUST}

\subsection{Nebular morphology}

\begin{figure}
\centerline{\includegraphics*[width=\linewidth]{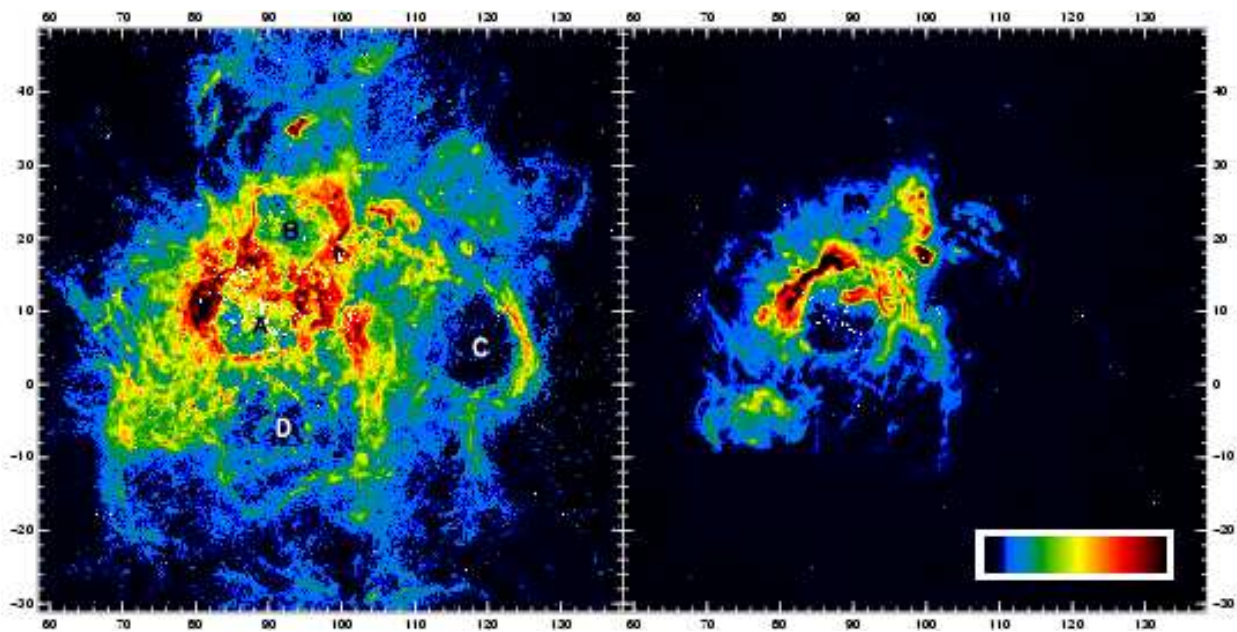}}
\caption{WFPC2 continuum-subtracted \sii\ (left) and \oiiir\ (right) images. 
Both images are displayed with a hyperbolic intensity scale in order to
show both low and high intensity structures. White pixels are used to block
regions with bright stars for the sake of clarity (continuum subtraction is
not perfect there due to the under-sampled nature of the WFPC2 PSF). The
larger extension of the emission in the \sii\ image as compared to that of
the \oiiir\ is not an artifact of the choice of displayed levels. On the
contrary, the dynamic range (defined as the ratio between maximum and
minimum displayed levels) for the image on the left is 100 while that for
the image on the right is 1\,000. Therefore, if both lines were emitted
from regions of similar size, one would expect the \oiiir\ region to appear
larger instead of smaller. The letters A to D indicate the location of the
four cavities described in the text.  The field is 80\arcsec\ on a side and
the orientation is the same as in the rest of the figures. See
Fig.~\ref{slosrmaps} for an explanation of the coordinate system.}
\label{wfpc2siioiiir}
\end{figure}

        The nebular structure of NGC 604 consists of a central bright,
high-excitation region surrounded by a dimmer low-excitation halo. The
central region is detected in all bright nebular lines while the halo can
be easily seen only in low-excitation lines such as \sii\ or \niir\ (see
Fig.~\ref{wfpc2siioiiir}). The high excitation region is composed of a
$16\arcsec\times 14\arcsec$ shell surrounding a central cavity (which we
will call A following the nomenclature of \citealt{Maiz00}) centered on
coordinates (90\arcsec,9\arcsec)\footnote{See Fig.~\ref{slosrmaps} for an
explanation of the coordinate system.}; two filaments that extend N/S along
$x=100\arcsec$ and $x=104\arcsec$, respectively; and a filled \hii\ region
centered at (76\arcsec,$-$5\arcsec).

\begin{figure}
\centerline{\includegraphics*[width=\linewidth]{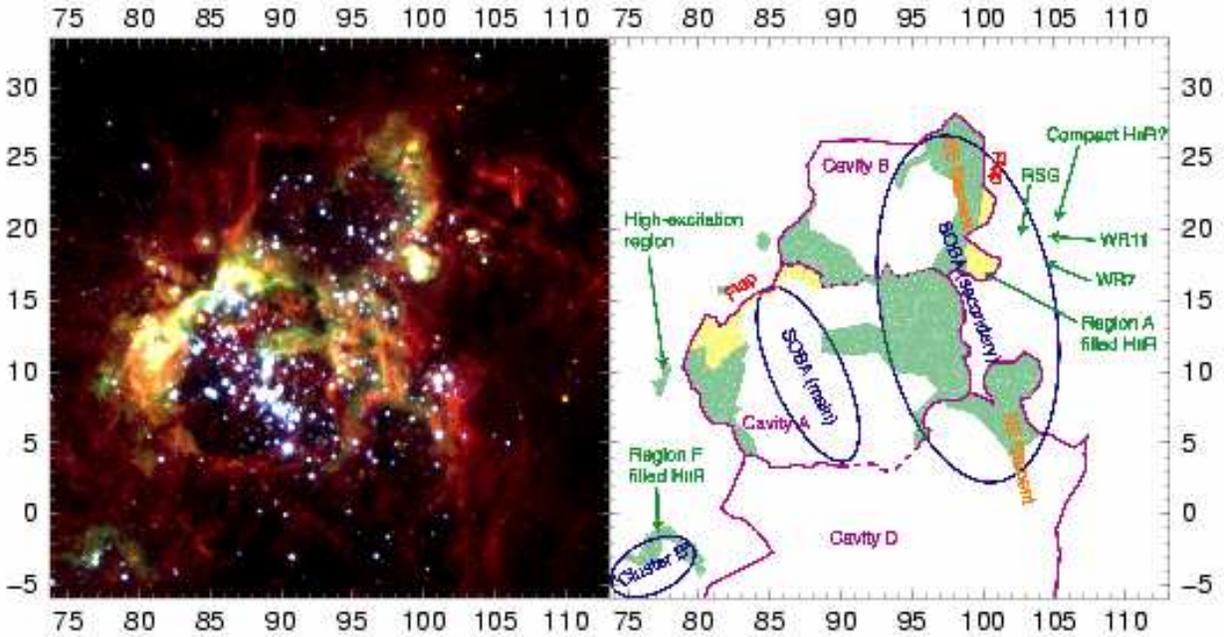}}
\caption{(left) Color mosaic of the central region of NGC 604 composed by assigning
F673N to the red channel, F555W and F656N to the green one, and F336W to the
blue one (the filters were not processed for continuum or line subtraction). 
The field is 39\arcsec\ on a side and the orientation is the same as in 
the rest of the figures. See Fig.~\ref{slosrmaps} for an explanation of the 
coordinate system. (right) Explanatory diagram for some of the structures 
discussed in the text. Violet is used to display cavity contours, with a dashed
line style when the boundary between cavities is uncertain (possibly due to 
superbubble bursting), and shading is
used to mark the high excitation parts of the nebula (\hii\ surfaces), with
yellow used for the regions with the highest intensity and green used for 
the rest.}
\label{colormosaic}
\end{figure}

        The SOBA has two loosely-defined components, a main one centered at
cavity A and elongated in the N-S direction (from approximately
(92\arcsec,2\arcsec) to (85\arcsec,18\arcsec), see Fig.~\ref{colormosaic})
and a secondary one, more dispersed and approximately cospatial with the
eastern part of the shell around cavity A and the two high-excitation
filaments located towards the E (in the range $x=92\arcsec-103\arcsec$ and
$y=2\arcsec-25\arcsec$, see Fig.~\ref{colormosaic}). Another more distant
group of bright stars can be seen at the location of the filled \hii\
region towards the NW (this group is called cluster B in
\citealt{Huntetal96}).

        The shell around cavity A is far from uniform. Towards the SW it is
very bright and very thin (likely unresolved) in \oiiir. Towards the E it
is of intermediate brightness, much thicker and less well defined, with a
bright intrusion extending towards the center of the cavity at (90\arcsec,
12\arcsec) \citep{TenTetal00}. Towards the N it is thin and well defined
again but dimmer. The two eastern filaments are also not uniform, with the
southern one brighter than the northern one. We want to stress that
the optical appearance of all the high excitation regions is consistent
with them being thin structures (thicknesses of $1-2$ pc), with an apparent
variable thickness created by different inclinations with respect to the
observer. As described in the introduction, this is the observed geometry
for well-studied galactic and extragalactic \hii\ regions. This effect is
seen, for example, in that the areas with the highest surface brightness
appear to be one-dimensional, as expected from orientation effects when the
geometry is such that we observe the surface edge on. Later on, we describe
other observational evidence that support this model for the structure of
the \hii\ gas in NGC 604.

        \citet{TenTetal00} explored the kinematics of the \hii\ gas in NGC
604 and found that the shell around cavity A can be characterized by a
single gaussian profile with an approximately constant velocity of $-255$
km s\m.  The \hii\ gas on cavity A itself needs to be described by at least
two kinematic components, one shifted towards the red and one towards the
blue. The red component is present everywhere and shows the spatial profile
of an expanding bubble while the blue one has a patchy coverage. The bright
intrusion at (90\arcsec, 12\arcsec) is part of the blue-shifted component.
\citet{TenTetal00} interpreted this kinematic profile as a sign that the
superbubble originally associated with cavity A had burst in the direction
towards us.

        Two other cavities are visible in the nebular-line images towards
the S and N; we will call them B and D, respectively, following the
previously established nomenclature (see Fig.~\ref{wfpc2siioiiir}). They
contain only a few massive stars each. Their kinematics were partially
mapped by \citet{TenTetal00} and they were found to be similar to that of
cavity A: the walls have single gaussian components with approximately
constant velocity while the cavity itself shows line-splitting indicative
of expansion. Given the low number of massive stars these two structures
contain and that they are apparently connected to the main cavity (there
are regions at the boundaries where only weak low-excitation nebular
emission can be seen), a likely origin for these two structures is that
they were also produced by the superbubble in cavity A bursting out into
the surrounding medium (in this case along the plane of the sky rather than
along the line of sight).

        The region towards the E of the $x=100\arcsec-104\arcsec$ filaments
is marked by a sharp drop in the excitation ratios (see
Fig.~\ref{wfpc2ratios}). A fourth cavity (C) is centered on
(119\arcsec,5\arcsec). It differs from the previous three in that the
intensity of the nebular lines in its central region is weaker and does not
show line splitting. Only a few stars are present inside and, judging from
their magnitudes, they are likely to be of late-O or early-B type at
most. All of this suggests that the fourth cavity is an older structure
which has been recently re-ionized by the current burst.

\begin{figure}
\centerline{\includegraphics*[width=\linewidth]{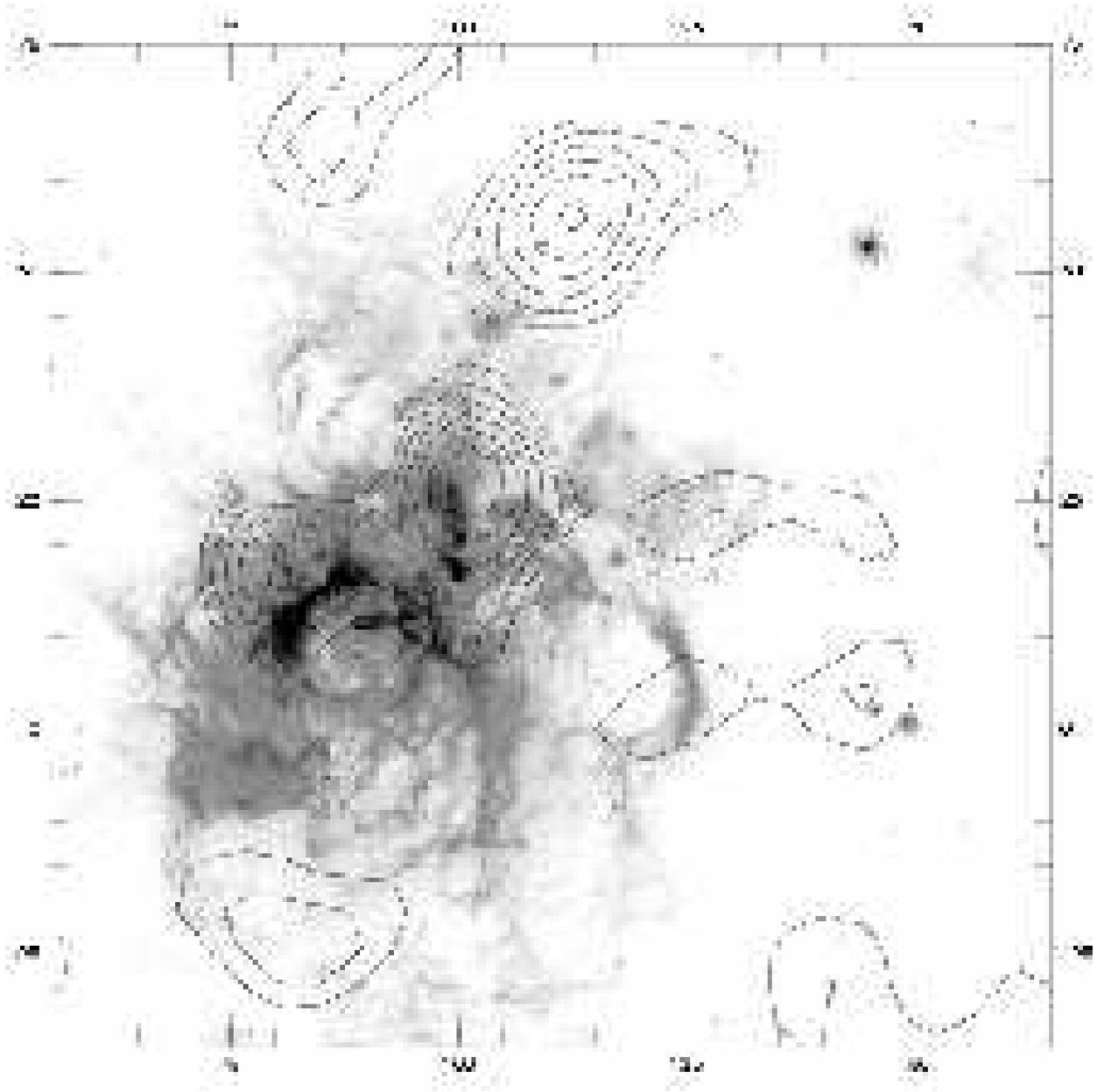}}
\caption{Contour diagram of the CO ($1\rightarrow 0$) emission in NGC 604 
(adapted from the data of \citealt{Engaetal03}) superimposed on an F656N
WFPC2 image (without continuum subtraction). The spatial resolution of the
contour diagram is 13\arcsec.  The field is 110\arcsec\ on a side and the
orientation is the same as in the rest of the figures. See
Fig.~\ref{slosrmaps} for an explanation of the coordinate system.}
\label{halfaco}
\end{figure}

        The distribution of the CO ($1\rightarrow 0$) emission in NGC 604
with respect to the \hii\ region is shown in Fig.~\ref{halfaco}. Two clouds
can be seen close to the SOBA. The largest one is centered on
(102\arcsec,27\arcsec) while the second one is centered on
(80\arcsec,17\arcsec). Farther away, a third cloud is centered on
(113\arcsec,51\arcsec) and a fourth, much less intense, on
(82\arcsec,-23\arcsec). Looking at Fig.~\ref{halfaco} and allowing for the
difference in spatial resolution between the two data sets, it can be seen
that all the bright areas of the \hii\ region are (a) located at the edges
of the first two molecular clouds and (b) oriented towards the ionizing
sources\footnote{That is, they are on the edges of the molecular clouds
that point towards the massive stars in the SOBA.}.  As we have already
mentioned, this is the same configuration observed in Galactic \hii\
regions and in more massive objects such as 30 Doradus and N11 and provides
further evidence towards the near-2D character of these areas. Therefore,
it appears to be a general configuration that the bright areas of \hii\
regions (excepting maybe the oldest and the youngest ones) originate in the
surface of molecular clouds directly exposed to massive young stars.

\begin{figure}
\centerline{\includegraphics*[width=\linewidth]{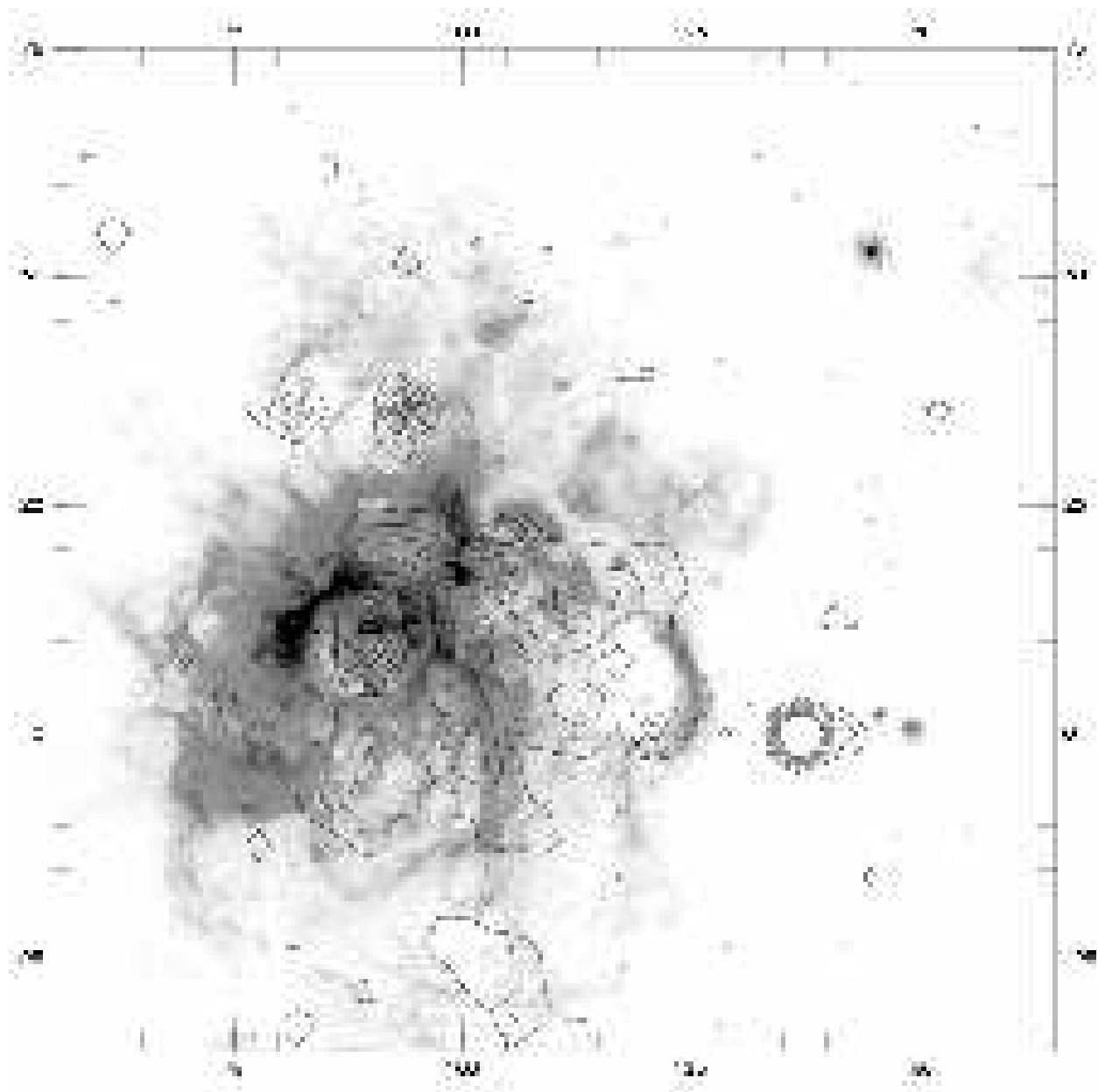}}
\caption{Contour diagram of the X-ray emission in NGC 604 superimposed on an 
F656N WFPC2 image (without continuum subtraction). The field and orientation is
the same as in Fig.~\ref{halfaco}.}
\label{halfax}
\end{figure}

	Diffuse X-rays are observed filling the four cavities in NGC 604 (see
Fig.~\ref{halfax}), with the highest intensity originating in the main
one. Two point sources are also detected: one is the SNR and the other one,
located at (137\arcsec,0\arcsec), is not associated with any of the other
sources described in this paper. The integrated X-ray emission from the
diffuse gas was modeled by \citet{Yangetal96}. The X-ray emission is very
soft, with a best-fitting Raymond-Smith model with a plasma temperature of
$1.3\times10^6$ K ($kT=0.12$ keV), giving a luminosity in the 0.5--2 keV
range of around $8\times10^{38}$ erg s$^{-1}$. These authors concluded that
the X-ray emission is most likely dominated by the contribution from a hot,
thin plasma energized by stellar winds. \citet{Cervetal02} predict a soft
X-ray emission of around $10^{34}$ erg s$^{-1}$ M$_{\odot}^{-1}$ for a MYC of 3
Myr, assuming average conditions. For a total stellar mass of around $10^5$
M$_\odot$ in the range 2--120 M$_\odot$ \citep{Yangetal96}, we conclude that
the observed X-ray luminosity is indeed consistent with a hot plasma
energized by a young massive cluster. As discussed by \citet{Cervetal02},
the contribution by individual stars of any kind at this age should be only
around $10^{30}$ erg s$^{-1}$ M$_{\odot}^{-1}$, which is negligible in
comparison with the contribution of the coronal gas. 

\subsection{Attenuation data}

\begin{figure}
\centerline{\includegraphics*[angle=90,width=\linewidth]{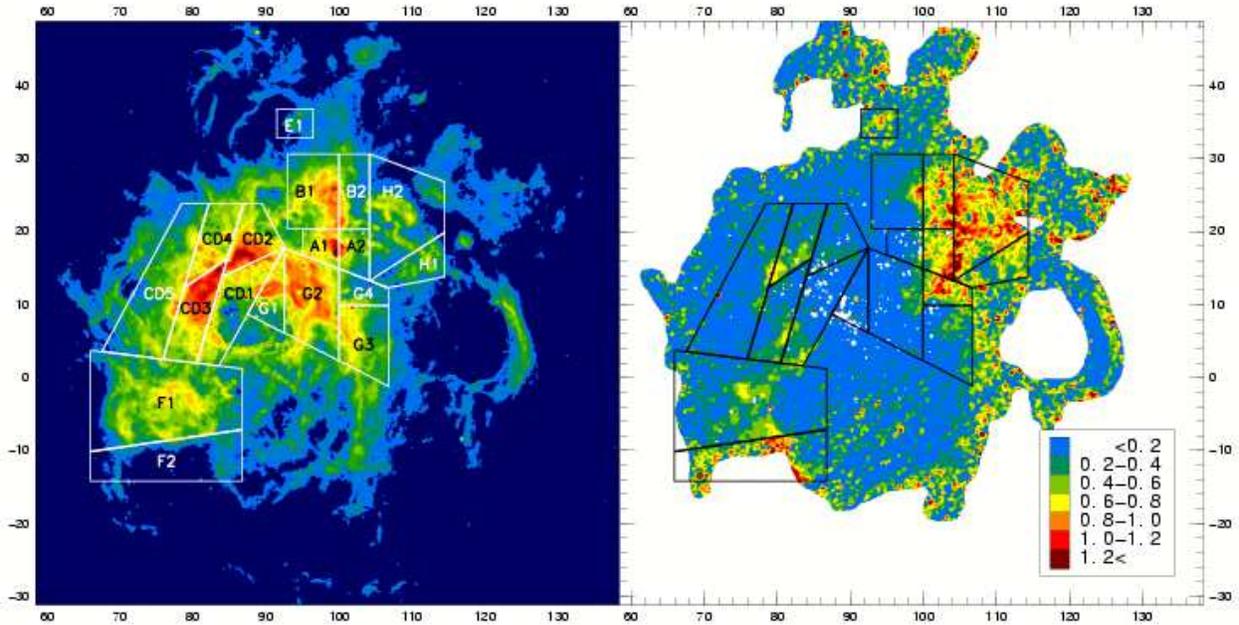}}
\caption{(Left) Continuum-subtracted \ha\ WFPC2 image of NGC 604. 
(Right) \taubal\ map of the
same region as derived from WFPC2 data smoothed with a $5\times 5$ WF pixels
box. Areas in white have been masked due to strong stellar contamination or low
signal-to-noise. The sub-regions used in this paper are shown.
The field is the same as the one in Fig.~\ref{wfpc2siioiiir}.
See Fig.~\ref{slosrmaps} for an explanation of the coordinate system.}
\label{wfpc2extinction}
\end{figure}

        We have smoothed our continuum-subtracted \ha\ and \hb\ images with
a $5\times 5$ pixel box, calculated their ratio, and applied
Eq.~\ref{tau2a} to obtain a map of \taubal, the optical depth at \ha\
measured from the ratio of the two Balmer lines. The result is shown in
Fig.~\ref{wfpc2extinction}, where we have masked out bright stars (as
previously described) and areas with low values of the S/N ratio. We stress
again the need for an accurate registering and flux calibration for such a
map to be a realistic representation of \taubal, topics that we have
covered in the previous section. Other aspects that require some words of
caution regarding Fig.~\ref{wfpc2extinction} are also detailed here:

\begin{itemize}
  \item An incorrect temperature can induce systematic differences in the value
        of \taubal\ (see Eq.~\ref{tau2} in the Appendix). For NGC 604 we use a 
        value of 8\,500 K based on the measurements of $8\,150\pm 150$ K for 
        $T_e$([\ion{O}{3}]) and $8\,600\pm 450$ K for $T_e$([\ion{N}{2}]) of
        \citet{Esteetal02} (see also \citealt{Diazetal87,GonDPere00}). 
        However, the temperature dependence is rather weak: a change from 
        8\,500 K to 10\,000 K (which is too large to be realistic for 
        NGC 604, see \citealt{Diazetal87,Esteetal02}) changes \taubal\ only from 
        0.245 to 0.273 for $F_{\alpha}/F_{\beta} = 3.2$. The dependence is 
        stronger for the case of \taurad\ (see Eq.~\ref{tau1}) but the 
        uncertainties introduced by a possible error in the temperature are not 
        large. A change from 8\,500 K to 10\,000 K would 
        change the overall value of \taubal\ for NGC 604 only from 0.54 to 0.63.
  \item The existence of scattered Balmer radiation from
        a dust cloud can yield localized values of 
        $F_{\alpha}/F_{\beta}$ lower than the canonical value corresponding to
        $\taubal=0$. Such a phenomenon has been observed in e.g. NGC 4214
        \citep{Maizetal98} and it is also present here in a few pixels in the
        central cavity. In those cases we have simply used $\taubal=0$.
\end{itemize}

        Our \taubal\ map provides us with a 0\farcs 5 (2 pc) resolution map
of the attenuation experienced by the \hii\ gas in NGC 604.  The problem of
recovering the flux lost by geometrical effects in the dust distribution
(as opposed to the flux lost by the total amount of dust present between
the source and the observer) can be divided into two parts: (a) the
patchiness of the dust geometry in the plane of the sky and (b) the
possibility that some of the dust may be interspersed among sources along
the same line of sight. (a) has the effect of underestimating the real \ha\
flux.  If we have observations with a spatial resolution comparable to the
scales in which the dust distribution varies, then we can correct for the
effects of dust by applying an extinction correction pixel by pixel. This
is the main motivation for producing a \taubal\ map with the highest
possible spatial resolution.

        The solution to (b) is more complicated. Suppose we have a thick
dust cloud located between sources A and B, both in the same line of sight
with A closer to the observer.  In that case, most or all of the Balmer
photons emitted from B will be absorbed in the cloud and our value of
\taubal\ will be only a measurement of the possible foreground extinction
experienced by A. From the point of view of \ha\ and \hb\ the existence of
B will be unknown to us and if we were to estimate the ionizing flux from
that information, we would miss the contribution from B completely. In a
general case, it can be shown that the effect of (b) is the same as that of
(a): to introduce an underestimation of the total ionizing flux. What can
we do to recover the lost flux? One solution is to use the thermal radio
continuum, which is unaffected by dust and can give us a better estimate of
the ionizing flux. Nevertheless, there are two considerations that affect
radio observations:

\begin{itemize}
  \item At the present time, radio observations of low-to-medium 
        emission-measure thermal sources such as extragalactic \hii\ regions 
        have lower spatial resolution than that available from optical 
        observations. The VLA can reach HST-like resolution but only for high 
        emission-measure sources. 
  \item It is possible to have contamination from non-thermal continuum radio 
        sources such as SNRs. This can be mitigated by observing at multiple 
        wavelengths but such a procedure can introduce large uncertainties if 
        the non-thermal contribution is large.
\end{itemize}

In addition, a fraction of the ionizing photons might simply be destroyed
by dust and/or might escape the region without contributing to the
ionization process. In our next paper we will characterize the massive
stellar population and will compare the predicted with the measured fluxes,
aiming to constrain the fraction of escaping photons.  

        In order to explore the relative importance of the non-uniform dust
distribution in the plane of the sky and along the line of sight we have
produced an $\approx 4\arcsec$ resolution map of \taurad, the real optical
depth at \ha\ as measured from the ratio of the thermal radio continuum to
\ha\ (see Eq.~\ref{tau1a} in the Appendix). The map is shown in
Fig.~\ref{radioextinction} along with the original radio image, the
resolution-degraded WFPC2 \ha\ image used to calculate \taurad, and a
\taubal\ map at the same resolution and scale as the \taurad\ one. We have
also integrated the fluxes along the polygonal regions shown in that same
figure, which correspond to the main \hii\ knots\footnote{The notation
follows \citet{ChurGoss99} joining C and D due to their proximity and
including G and H as two new knots. We also include the sum: (a) of the 7
regions, (b) of all the integrated area, and (c) of all the integrated area
without knot E (the SNR).}, and have computed the corresponding values of
\taurad\ and \taubal\ (see Table~\ref{integr1}). In that same table we
include three other quantities. The first one, \taubalp, is the value of
\taubal\ calculated by correcting for extinction pixel by pixel in the
region using the full-resolution WFPC2 images, i.e. the weighted value we
recover by using the spatial information in the plane of the sky up to the
resolution provided by WFPC2. One expects that $\taubal < \taubalp <
\taurad$, since \taubalp\ does not correct for structures present in the
dust at smaller spatial scales and along the line of sight, and that is
indeed the case for all entries in Table~\ref{integr1}. The other two
quantities are the values of \taucov\ and $\gamma$ corresponding to
\taurad\ and \taubal\ applying the patchy foreground model of the Appendix
and using the one-to-one mapping that can be derived from
Fig.~\ref{extplot}.

\begin{deluxetable}{lrrrcrrcrrrr}
\tablecaption{Integrated values for A to H regions.\label{integr1}}
\tabletypesize{\scriptsize}
\tablewidth{0pt}
\tablehead{\colhead{Apert.} & \colhead{Area\tablenotemark{a}} & \colhead{$F_{\rm rad}$\tablenotemark{b}} & \colhead{$F_{\alpha}$\tablenotemark{c}} & \colhead{$F_{\rm rad}/F_{\alpha}$\tablenotemark{d}} & \colhead{\taurad\tablenotemark{e}} & \colhead{$F_{\beta}$\tablenotemark{c}} & \colhead{$F_{\alpha}/F_{\beta}$} & \colhead{\taubal\tablenotemark{e}} & \colhead{$\tau^{\prime}_{\rm Bal}$\tablenotemark{f}} & 
\colhead{\taucov\tablenotemark{g}} & \colhead{$\gamma$\tablenotemark{g}}}
\startdata
A               &   48.80 &   3.62 &   13.73 &  0.264 &  0.92 &    4.12 &  3.330 &  0.34 &  0.47 &  2.02 & 0.695 \\
B               &  116.39 &   6.55 &   20.96 &  0.313 &  1.09 &    6.24 &  3.357 &  0.36 &  0.45 &  2.27 & 0.741 \\
CD              &  347.47 &  17.92 &   87.71 &  0.204 &  0.67 &   28.13 &  3.118 &  0.18 &  0.25 &  2.17 & 0.549 \\
E               &   93.96 &   0.67 &    3.48 &  0.192 &  0.61 &    1.07 &  3.258 &  0.29 &  0.51 &  1.48 & 0.588 \\
F               &  349.04 &   4.82 &   29.35 &  0.164 &  0.45 &    9.05 &  3.244 &  0.28 &  0.40 &  1.03 & 0.561 \\
G               &  208.20 &   6.90 &   47.22 &  0.146 &  0.33 &   15.71 &  3.006 &  0.09 &  0.21 &  1.89 & 0.333 \\
H               &  161.82 &   3.29 &   11.09 &  0.297 &  1.04 &    2.92 &  3.798 &  0.66 &  0.80 &  1.52 & 0.827 \\
A-H             & 1325.68 &  43.78 &  213.55 &  0.205 &  0.67 &   67.25 &  3.176 &  0.23 &  0.34 &  1.92 & 0.572 \\
NGC 604         & 6120.94 &  56.77 &  314.60 &  0.180 &  0.54 &   99.17 &  3.172 &  0.22 &  0.42 &  1.59 & 0.525 \\
NGC 604 (no E)  & 6026.98 &  56.10 &  311.12 &  0.180 &  0.54 &   98.10 &  3.172 &  0.22 &  0.41 &  1.59 & 0.525 \\
\enddata
\tablenotetext{a}{In square arcseconds.}
\tablenotetext{b}{In mJy.}
\tablenotetext{c}{In $10^{-13}$ erg s\m\ cm\mm .}
\tablenotetext{d}{In $10^{13}$ mJy erg\m\ s cm$^2$.}
\tablenotetext{e}{Calculated from previous column (low resolution data).}
\tablenotetext{f}{Calculated by correcting for extinction pixel by pixel in the high resolution data.}
\tablenotetext{g}{Calculated from \taurad\ and \taubal, see Appendix and Fig.~\ref{extplot}.}
\tablecomments{The two largest uncertainty sources for the optical depths are 
               the absolute flux calibration and the assumed temperature. See 
	       the discussion in subsections 2.2 (see Table~\ref{wfpc2cal}) and 3.2, respectively.}
\end{deluxetable}

        In order to make better use of the high-resolution optical data, we
have subdivided the A to H regions into sub-regions and calculated the
corresponding values of \taubal\ and \taubalp.  The results are shown in
Table~\ref{integr2}, where the entry ``diffuse'' refers to the totality of
the integrated area except the previous line, which is the sum of all
sub-regions. Note that the value of \taubalp\ for the ``diffuse'' region is
likely to be an overestimation of the real value caused by the low S/N
ratio of the Balmer ratio for the outer regions of NGC 604 because the
logarithm in Eq.~\ref{tau2a} introduces a bias by giving more weight to the
pixels where $F_\beta$ is close to zero due to noise fluctuations.

\begin{deluxetable}{lrrcrrccc}
\tablecaption{Integrated values for A1 to H2 subregions.\label{integr2}}
\tabletypesize{\scriptsize}
\tablewidth{0pt}
\tablehead{\colhead{Apert.} & \colhead{Area\tablenotemark{a}} & \colhead{$F_{\alpha}$\tablenotemark{b}} & \colhead{$F_{\alpha}/F_{\beta}$} & \colhead{\taubal\tablenotemark{c}} & \colhead{$\tau^{\prime}_{\rm Bal}$\tablenotemark{d}} & \colhead{ratio 1\tablenotemark{e}} & \colhead{ratio 2\tablenotemark{f}} & \colhead{ratio 3\tablenotemark{g}}} 
\startdata
A1              &   22.51 &    9.88 &  3.203 &  0.25 &  0.33 &  2.63 &  0.094 &  0.090 \\
A2              &   26.26 &    4.25 &  3.757 &  0.63 &  0.73 &  2.42 &  0.159 &  0.135 \\
B1              &   73.13 &   16.16 &  3.253 &  0.28 &  0.36 &  2.07 &  0.121 &  0.104 \\
B2              &   43.26 &    5.10 &  3.712 &  0.61 &  0.69 &  1.74 &  0.196 &  0.151 \\
CD1             &   69.39 &   14.03 &  2.996 &  0.09 &  0.25 &  1.75 &  0.136 &  0.117 \\
CD2             &   44.07 &   19.80 &  3.103 &  0.17 &  0.23 &  2.49 &  0.088 &  0.081 \\
CD3             &   55.82 &   28.77 &  3.145 &  0.20 &  0.22 &  2.10 &  0.105 &  0.099 \\
CD4             &   43.94 &    9.83 &  3.307 &  0.32 &  0.38 &  1.78 &  0.149 &  0.132 \\
CD5             &  129.45 &   14.38 &  3.108 &  0.17 &  0.24 &  1.77 &  0.155 &  0.130 \\
E1              &   20.91 &    0.87 &  3.414 &  0.40 &  0.57 &  1.48 &  0.553 &  0.218 \\
F1              &  230.97 &   25.32 &  3.209 &  0.25 &  0.31 &  1.82 &  0.156 &  0.121 \\
F2              &  116.01 &    3.84 &  3.500 &  0.46 &  0.81 &  1.50 &  0.240 &  0.170 \\
G1              &   29.65 &    7.21 &  2.877 &  0.00 &  0.21 &  1.93 &  0.118 &  0.102 \\
G2              &   90.78 &   29.32 &  3.002 &  0.09 &  0.16 &  1.70 &  0.122 &  0.106 \\
G3              &   63.25 &    9.48 &  2.945 &  0.04 &  0.21 &  1.15 &  0.205 &  0.155 \\
G4              &   25.17 &    2.15 &  3.598 &  0.53 &  0.68 &  1.27 &  0.229 &  0.172 \\
H1              &   36.67 &    2.77 &  3.510 &  0.47 &  0.56 &  1.16 &  0.245 &  0.166 \\
H2              &  124.68 &    8.54 &  3.925 &  0.74 &  0.87 &  1.48 &  0.234 &  0.163 \\
A1-H2           & 1245.92 &  211.70 &  3.175 &  0.23 &  0.34 &  1.90 &  0.141 &  0.116 \\
diffuse         & 4875.02 &  102.90 &  3.168 &  0.22 &  0.56 &  1.32 &  0.266 &  0.194 \\
NGC 604         & 6120.94 &  314.60 &  3.173 &  0.22 &  0.42 &  1.71 &  0.182 &  0.142 \\
NGC 604 (no E1) & 6100.03 &  313.73 &  3.172 &  0.22 &  0.42 &  1.71 &  0.181 &  0.142 \\
\enddata
\tablenotetext{a}{In square arcseconds.}
\tablenotetext{b}{In $10^{-13}$ erg s\m\ cm\mm .}
\tablenotetext{c}{Calculated from previous column.}
\tablenotetext{d}{Calculated by correcting for extinction pixel by pixel.}
\tablenotetext{e}{[O\,{\sc iii}]~$\lambda$4959/H$\beta$.} 
\tablenotetext{f}{[S\,{\sc ii}]~$\lambda$6717+6731/H$\alpha$.} 
\tablenotetext{g}{[N\,{\sc ii}]~$\lambda$6584/H$\alpha$.}
\tablecomments{The two largest uncertainty sources for the optical depths are 
               the absolute flux calibration and the assumed temperature. See 
	       the discussion in subsections 2.2 (see Table~\ref{wfpc2cal}) and 3.2, respectively}
\end{deluxetable}

        We end this subsection by pointing out that the values listed in
Tables~\ref{integr1}~and~\ref{integr2} for the entry ``NGC 604'' refer to
the area (a) where nebulosity can be easily detected in \ha\ (b) that is
covered by all the WFPC2 exposures in programs 5237, 5773, and 9134. In
order to check that we were not leaving out any outlying regions with a
significant flux contribution, we compared our total value for the \ha\
flux with the one measured by \citet{Boscetal02} using a larger area. Their
value, \scipm{3.1}{0.4}{-11} erg cm\mm\ s\m, is in excellent agreement with
ours.

\subsection{Results}

\begin{figure}
\centerline{\includegraphics*[angle=90,width=\linewidth]{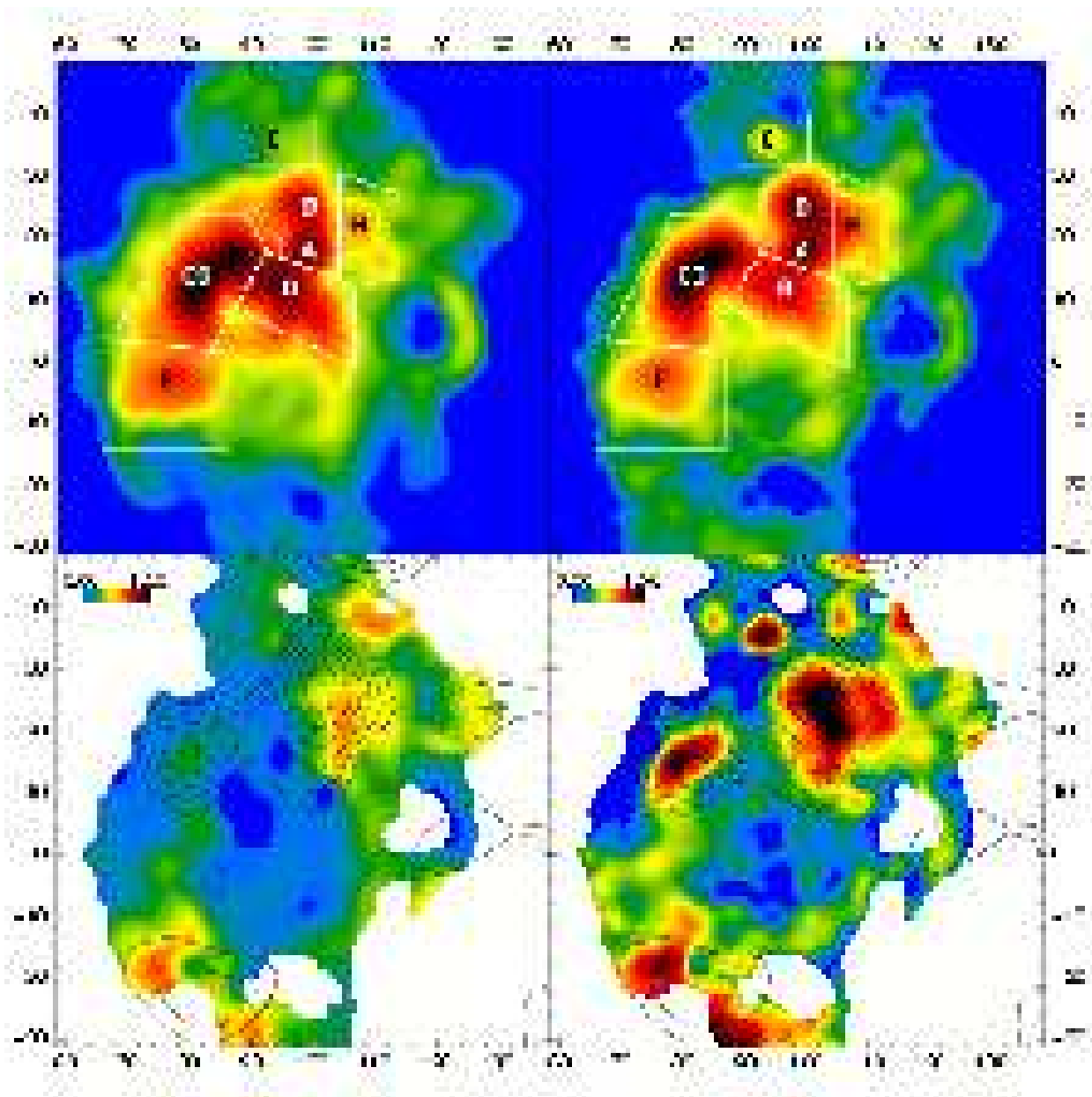}}
\caption{(Upper left) Low ($\approx 4\arcsec$) resolution continuum-subtracted 
\ha\ image of NGC 604. (Upper right) 8.4 GHz radio continuum image of NGC 604
at the same resolution. (Lower left) \taubal\ map of the same region at that 
resolution. (Lower right) \taurad\ map of the same region at that resolution. 
The regions used in this paper are shown in the upper two panels.
Areas in white in the two lower panels have been masked due to low 
signal-to-noise. Contours from the CO data shown in Fig.~\ref{halfaco} are 
plotted in the lower two panels. The field is the same as the one in 
Fig.~\ref{wfpc2siioiiir}.  See Fig.~\ref{slosrmaps} for an explanation of the 
coordinate system.}
\label{radioextinction}
\end{figure}

        Four attenuation maxima are present in the \taurad\ map of
Fig.~\ref{radioextinction}, three in the upper half of the map centered around
(104\arcsec,24\arcsec), (80\arcsec,15\arcsec), and (93\arcsec,35\arcsec),
respectively, and another one more extended near the bottom of the map. The
third maximum is not really a high attenuation region: it is the previously
mentioned SNR, for which Eq.~\ref{tau1a} does not provide an accurate
measurement of \taurad\ given the non-thermal character of the radio flux, and
will not be discussed any further here. The other three, however, coincide with 
the positions of three of the CO ($1\rightarrow 0$) 
maxima in Fig.~\ref{halfaco}. In general, once
we account for the SNR and the difference in resolution, there is a very good
correlation between \taurad\ and CO ($1\rightarrow 0$) intensity.

        If we try the same comparison using the \taubal\ map instead of
\taurad, we find that the correlation is also present but that it is not as
strong. The main culprit is the attenuation maximum at (80\arcsec,15\arcsec),
which is much weaker in \taubal\ than in \taurad. It is also clear from 
Fig.~\ref{radioextinction} that $\taurad \ge \taubal$ everywhere (except for a
few weak areas where resolution and scattering effects may be important), as
expected from the discussion in the previous subsection.

        The first attenuation maximum corresponds to the main molecular cloud
of NGC 604 (cloud 1 of \citealt{Vialetal92}, cloud 2 of \citealt{WilsScov92})
and extends over region H and part of regions A, B, and G (more
specifically, over sub-regions A2, B2, G4, H1, and H2).
The right panel of Fig.~\ref{cuts1} shows the correlation between \taurad\ and 
CO($1\rightarrow 0$) intensity in this region. The
brightest regions in \ha\ are located in A and B, which span the SE bright
filament. A strong E-W attenuation gradient is visible on those regions, as it
can be seen comparing A1 with A2 and B1 with B2 (also G2 with G4) in 
Fig.~\ref{wfpc2extinction} and in the cuts in Fig.~\ref{cuts2} (note how in 
the first plot the attenuation gradient coincides with the \ha\ intensity 
maximum at the center of region A). The existence of this gradient could in
principle justify the fact that $\taubal\approx \taurad/3$ for both A and B by
indicating that a patchy foreground screen is responsible for the difference.
However, \taubalp/\taurad\ is 0.51 for A and 0.41 for B, so either the 
patchiness extends to scales below 2 pc or some Balmer emission is completely 
hidden from view. In this context, it is interesting to note that the \ha\ and
radio peaks are nearly coincident for A but that the radio peak is displaced
towards the SE by $\approx 1\arcsec$ with respect to the \ha\ peak for B, the
region of the two with the lowest value of \taubalp/\taurad. 
This suggests that the \hii\ gas 
is located along the surface of the main molecular cloud and that this cloud
creates a high-obscuration ``flap'' (seen in Fig.~\ref{colormosaic} running
from (101\arcsec,21\arcsec) to (100\arcsec,27\arcsec)) that absorbs most or all
of the Balmer photons located behind, letting only the radio continuum photons
traverse it. The interpretation of that intensity drop as such a flap would be
consistent with the observed values of the optical depth and, as we will see
later, may not be the only such structure in NGC 604.

        Region A, on the other hand, appears to be dominated by a
single, compact, barely-resolved \hii\ region. Our measured value of 
\taubalp/\taurad\ is more likely to be explained by our lack of better
spatial resolution. Farther towards the E into the main
molecular cloud, region H shows high attenuation over most of its extension 
but does not include very bright areas. Its values for \taurad\ and \taubalp\ 
are quite similar, an indication that a simple patchy foreground screen 
provides an accurate description of the \hii\ region without the need to 
include much highly-obscured \hii\ gas or unresolved dust clouds.

        The second attenuation maximum corresponds to a molecular cloud that was
outside the region studied in CO by \citet{Vialetal92} and \citet{WilsScov92} 
but that is well defined in the \citet{Engaetal03} map. This region also
corresponds to the SW quadrant of the main cavity, where the high-excitation 
shell is brightest, and is covered by our low-resolution region CD. We note
the following:

\begin{itemize}
  \item \taubal, \taubalp, and \taurad\ here are lower than in regions A, B, 
        or H, indicating an overall lower importance of attenuation.
  \item $\taubalp/\taurad =0.39$, a value similar to that of region B, 
        suggesting the existence of hidden \hii\ gas. 
  \item The values of \taubal\ ($\approx 0.2$) for the sub-regions
        CD2 and CD3 that correspond to the brightest regions (the high 
        excitation shell) are almost identical to the overall value of \taubal\ 
        for the whole CD region. 
  \item A sharp intensity drop in all emission lines is seen between sub-regions
        CD2 and CD3 and subregion CD4 (sub-regions were 
        specifically chosen to follow this boundary), as it can be seen in
        Figs.~\ref{colormosaic}~and~\ref{cuts2}.
\end{itemize}
        
        All of the above point towards the existence of a more or less
uniform foreground extinction with $\taubal\approx 0.2$ combined with 
another ``flap'' that is covering the central part of the near-edge-on
high-excitation region, almost dividing it into two to the point of making it
appear as two separate knots (C and D) in the low-resolution \ha\ data
\citep{ChurGoss99}. Can we test this hypothesis any further? Yes, there are
three more pieces of evidence that point in the same direction. First, the radio
peaks are displaced by $\approx 2\arcsec$ towards the SW with respect to the
low-resolution \ha\ peaks, which is exactly what would be expected if the flap
detectable in the high-resolution \ha\ images was occulting a flux similar to
that in sub-regions CD2 and CD3. Second, the value of \taubal\ for subregion CD4
is indeed higher than that for CD2 and CD3 but nowhere near the value required
to raise the total value of \taubal\ for all of CD to 0.67. This is also seen in
Fig.~\ref{cuts2}, where the sharp drop in \ha\ intensity as one moves towards 
the left (W) is accompanied only by a slow increase in \taubal\ for 
$y=16\farcs 6$ (left panel) and by a more abrupt (but still insufficient to 
raise the total \taubal\ to 0.67) one at $y=13\farcs 4$ (right panel). 
What is likely happening here is that we are seeing in the optical is 
ionized gas near the front side of the attenuation flap, likely
affected by it, but not the material behind it, which is obscured to the point
that we can only see it in radio continuum. Third, if we apply the model
described in the Appendix, we obtain $\taucov=2.17$ and $\gamma=0.549$. The
value of $\gamma$ appears to be consistent with the observed morphology but
\taucov\ is too small (if it were really that low, we would detect a much
higher value of \taubal\ at the location of the flap). However, one should bear 
in mind that the model assumes no attenuation for the uncovered region and, as 
we have seen, the low-attenuation
part of CD has $\taubal\approx 0.2$. If we add an additional uniform screen
that affects all the ionized gas with $\taubal= 0.18$, all the curves in
Fig.~\ref{extplot} are displaced towards the top and we end up with a $\gamma$
of approximately 0.5 and an arbitrarily large \taucov. In this sense, the
values of \taucov\ provided in Table~\ref{integr2} should be understood only as 
a lower limit (on the other hand, it is easy to see from Fig.~\ref{extplot} 
that the values of $\gamma$ are not so strongly affected by the vertical
displacement induced by a uniform foreground screen if its optical depth is not
too large). We can conclude that all available evidence suggests the
existence of a high-attenuation flap that covers a significant fraction of the 
SW high-excitation shell.

        The last attenuation maximum, located towards the N of subregion F1,
corresponds to a weak CO ($1\rightarrow 0$) maximum. In this case, the effect
on the \hii\ gas in region F is not as important ($\taurad = 0.45$) and 
\taubalp\ is very close to \taurad, so a patchy foreground screen with a low
value of \taucov\ provides a good fit to the observed properties (this can be
seen also in the proximity to the diagonal (\taubal=\taurad) of F in 
Fig.~\ref{extplot}). The
last of the regions, G, is located along the eastern edge of
the main cavity and also includes the NE high-excitation filament. Most of it
experiences low attenuation, with the only exception of G4, which covers the
northern tip of the attenuation feature of the main molecular cloud. All of
this results in the lowest values of \taurad, \taubal, and \taubalp\ among all
regions. Note also how the simple patchy foreground model of
the Appendix correctly characterizes what is seen in the high resolution data:
the value of \taucov\ is similar to that of its neighbors, A and H, but
$\gamma$ is much lower since attenuation affects significantly only G4. 

        On the issue of the relative location of the CO clouds with respect to
the extinction experienced by different sources it is interesting to point out
the recent result by \citet{Bluhetal03}. Those authors used FUSE to try to
detect H$_2$ absorption towards NGC 604 but they failed to do so. The
morphology described in this paper easily explains the reasons for that
non-detection: it is not that there is no molecular gas in NGC 604 but rather
that there is very little in front of most of the UV-bright stars. Note that
even if there were many massive stars embedded in the CO cloud it would be hard
to detect the H$_2$ signature in the integrated FUV spectrum, since those stars
would be too extinguished in that wavelength range and most of the detected FUV
photons would come from the unextinguished stars. In order to detect a strong
FUV H$_2$ signature in a integrated spectrum, it is required that the molecular 
gas covers most or all of the massive stars. 

\section{DENSITY AND EXCITATION}

\subsection{Density maps}

        We show in Fig.~\ref{slosrmaps} the maps of the electron density ratio 
\siil/\siir\ derived from the long-slit data, along with a synthetic \ha\
map obtained in the same way. Given that low S/N data have large uncertainties
that render the values of the density ratios meaningless, only those points
where the uncertainty in \siil /\siir\ (as derived from the fitting to the
spectra) is less than 0.08 are shown. In order to extract as much information as
possible, two maps and one table are shown at different spatial resolutions.
Lower spatial resolutions allow us to extend the maps to larger areas,
increasing the S/N by smoothing over more pixels. 

        Most of NGC 604 has values of $\siil /\siir > 1.30$, which corresponds
to the low-density regime $\dens \lesssim 130$ cm\mmm\ \citep{Castetal92}. 
The low density areas include the central cavity, most of the surrounding shell,
and the blue-shifted \ha\ knot. Only a few small regions of the main shell show 
values of $\siil /\siir \approx 1.30$ in the high-spatial resolution map. The
fact that most of the bright regions of the shell are not especially dense is
another point in favor of the description of these regions as 2-D surfaces since
it implies that their high surface brightness is caused by large column 
densities (explained by near-edge-on orientations of the surfaces) and not by 
high densities\footnote{Note that we are unable to calculate detailed filling 
factors for the gas in NGC 604 using a combination of the emission measure and
the density since we can only provide an upper bound to the density.}.

        Three regions are characterized by having a distinct high density in
Fig.~\ref{slosrmaps}. The first one is the compact \ha\ knot in region A 
centered at (100\arcsec,17\farcs3). Its high density is shown in the two maps in
Fig.~\ref{slosrmaps} and also in the left panel of Fig.~\ref{cuts1}. There we can
see that the density is well correlated with the \ha\ intensity in the 5\arcsec\
around the knot, reaching a maximum value around 250 cm\mmm. Compare this with
the \ha-bright southern part of the main shell (around $x=87.5\arcsec$ in the 
same plot), which shows lower densities and a much poorer correlation between
intensity and density.

        The second high-density region is located at the southernmost (top) 
edge of the maps around (105\arcsec,20\farcs 9). 
The region is dense enough to appear in
yellow in the bottom table of Fig.~\ref{slosrmaps}. We also show in the right
panel of Fig.~\ref{cuts1} the density as a function of the $y$ coordinate, with 
a maximum value around 360 cm\mmm. The density gradient in that plot shows a 
good correlation with \taurad\, which itself increases as one goes into the 
largest molecular cloud in NGC 604. The higher-resolution \taubal\ map shows
that this specific region is highly extinguished, which leads to the
possibility that there may be a barely-visible compact \hii\ region. This area
is adjacent to the location of a stellar group centered at 
(103\arcsec,19\farcs 5) that contains a WR/Of star (WR11, \citealt{Drisetal93}) 
and a RSG \citep{TerEetal96}. This leads to the interesting possibility of
having a compact \hii\ region, a WR star, and a RSG within 4\arcsec\ (16 pc) of
each other, which is quite surprising, given that each of these objects
represents a different stage in the evolution of massive stars separated from
the rest by several Myr. It should be pointed out, however, that the current 
data does not exclude the possibility of a chance alignment between the RSG and
the other two objects. Furthermore, if the star with \heii\ excess (which is
really what \citealt{Drisetal93} detected) turned out
to be an Of star instead of a WR, there would not  necessarily be an age 
discrepancy with the compact \hii\ region. The situation would be similar to
that of knot 2 in 30 Doradus, where an early-O star is partially embedded
in a molecular cloud adjacent to the main stellar cluster \citep{Walbetal02a}.

        The third high-density region is located around 
(105\arcsec,17\farcs 3), 4\arcsec\ to the N of the second one. It can be seen in
the third panel of Fig.~\ref{slosrmaps} and as a secondary maximum in the right
panel of Fig.~\ref{cuts1}. This region corresponds to the location of another
WR/Of star, WR7 \citep{Drisetal93} and is further discussed in the next
subsection.

\subsection{Excitation structure}

        We show in Fig.~\ref{wfpc2ratios} maps of the three excitation
ratios derived from the WFPC2 data, \oiiir/\hb, \sii/\ha, and \niir/\ha. 
As we did for the \ha/\hb\ map, we first smoothed with a $5\times 5$ pixel box
the continuum-subtracted emission-line images, calculated their ratios, and
plotted the result with the low S/N ratio areas and bright stars masked out.
Inspection of emission line diagnostic diagrams (Fig.~\ref{diagn}) reveals two 
types of ionization structures: (a) a main central ionization structure (CIS) 
produced by the central SOBA, and (b) a number of secondary ionized structures 
(SISs) localized outside the main shell and energized by nearby small 
stellar groups.

        The CIS is dominated by the main shell (shell A) centered at
(90\arcsec,9\arcsec). The upper panel in Fig.~\ref{CIS-shell} shows the radial 
distribution of the CIS 
\ha\ flux (in log scale), where it can be seen that the general 
appearance is that of an empty shell integrated along the line of sight, 
with an inner radius of 8\arcsec, plus an extended low-surface-brightness halo. 
The overall appearance of the CIS is that of an inner 
$\approx 20\arcsec$ radius high excitation zone surrounded by an outer larger 
low excitation halo. The high excitation zone is relatively symmetrical about 
the main shell. The low excitation halo is asymmetrical; it extends out to 
$\approx 45\arcsec$ towards the North, East and South, but it is significantly
less important towards the West. Figure~\ref{diagn} shows the
diagnostic diagram for four quadrants and a difference is readily apparent
there in one of the four cardinal directions:
the W$\pm 45\arcdeg$ diagram has all its pixels in the high to intermediate 
excitation regime, $\log(\oiiir/\hb)>-0.2$, indicating
that the nebula is density bounded towards the West.

The lower panel in Fig.~\ref{CIS-shell} shows the log(\oiiir/\hb) vs. 
log(\sii/\ha) diagnostic diagram of the CIS. This diagram has been constructed
from the pixel-by-pixel diagnostic diagram of the points belonging to the CIS,
and where each square represents the density of individual points located
in that part of the diagram. The contours give the density of points in log 
scale, while color is used to code the average radial distance of those points
to the CIS center. We see that those points closer to the center (blue 
colors) have high excitation, while those further away (red) tend to be of 
lower excitation. Thus, this diagnostic diagram traces the overall ionization 
structure of the CIS, as expected for a simple shell+halo structure. We have 
run a series of models similar to those described in the simple approach 
followed by \citet{GonDPere00}. The density distribution used is taken as the 
azimuthal average rms density distribution as obtained from the \ha\ flux, and 
converted to actual density via the filling factor $\phi$ (taken as a 
parameter). For a range of ages between 1 and 5 Myr, and $\phi=0.1,0.01,0.001$, 
we have used Cloudy \citep{Ferl97} with the same input SEDs as in 
\citet{GonDPere00}. The output radial distribution of emission line fluxes are 
integrated along the line of sight, assuming spherical symmetry.
Only those models in the age range 2.75$-$3.0 Myr and $\phi=0.1$ fall close 
enough to the CIS ionization structure (see Fig.~\ref{CIS-shell}).
We notice that: (i) as concluded by \citet{GonDPere00}, the age of the 
cluster ionizing the CIS is 2.5$-$3 Myr; and (ii) simple spherically symmetric 
models produce a well defined line in the diagnostic diagram, while the CIS 
points in NGC 604 are distributed along a
thick region in the ionization structure. This spread in NGC 604 is 
produced by inhomogeneities in the detailed structure, with each 
radial direction from the center having a different actual density distribution.
An additional cause for the spread 
is the fact that the ionizing stars are not all located at the center of the 
CIS, i.e., NGC 604 is ionized by a SOBA and not by a SSC.

        Outside the main shell a number of SISs can be identified by their clear 
footprints on the diagnostic diagrams; they are ionized by a few or even just a 
single massive star. They are as follows.

\begin{itemize}
  \item The filled \hii\ region in region F, centered at (76\arcsec,5\arcsec)
        with a radius of 6\arcsec. The ionizing source consists of a small 
        group that produces a maximum $\log(\oiiir/\hb)=0.4$. The extinction is 
        low, with $\taurad = 0.45$.
  \item The region around WR7, located at (104\arcsec,18\arcsec) with a 
        radius of 1\farcs 5. It is ionized by a single WR/Of star, identified 
        as WR7 by \citet{Drisetal93}, and reaches a high excitation value of 
        $\log(\oiiir/\hb)=0.6$. This star is located in a ridge where 
        \taubal\ grows rapidly from 0.7 to 1.2. As mentioned in the previous
        subsection, it also corresponds to a region of high density as measured
        from the \siil/\siir\ ratio, possibly a compact \hii\ region. If the
        star and the compact \hii\ region are physically associated, the
        star should be an Of rather than a Wolf-Rayet.
  \item The filled \hii\ region in region A, centered at
        (100\arcsec,17\arcsec) and with a radius of 2\farcs 5, is the brightest 
        and more compact within NGC 604. Ionized by a handful of UV-bright 
        stars, it presents the highest ionization level, with up to 
        $\log(\oiiir/\hb)=0.75$. Located at the edge of the main molecular
        cloud, it has a value of \taurad\ of 0.92.
  \item The high-intensity \hii\ gas in region B is elongated along the SN 
        direction and just south of the filled \hii\ region in region A. It
        has an approximate size of $6\arcsec\times9\arcsec$ and its ionization 
        level is not very high, $\log(\oiiir/\hb)=0.35-0.40$, except in an 
        unresolved high excitation knot at (100\arcsec,26\arcsec) where it reaches 
        0.6.
  \item Region E corresponds to the SNR described by \citet{DOdoetal80} and it
        is centered at (94\arcsec,35\arcsec) with a radius of 1\farcs 5.
        Its ionization footprint is conspicuous only in the ratio 
        \sii/\ha, with logarithmic values between $-0.2$ and 0.0, with all
        other points in NGC 604 having values smaller that $-0.5$ for the log 
        of this ratio. The excitation ratio \oiiir/\hb\ has rather low 
        values (logarithm between $-0.3$ and 0.2) except for the SE part of the
        SNR, where an \oiiir-bright knot raises the value to 
        $\log(\oiiir/\hb)=0.4-0.5$. Contamination from the SNR hampers the
        measurement of the extinction, but an analysis of the surroundings
        indicate a low value. It is interesting to point out that the effect 
        of the SNR on the global emission-line ratios for NGC 604 is
        very small, as evidenced by comparing the last two lines in either
        Table~\ref{integr1} or Table~\ref{integr2}. It would be impossible
        to detect its presence using spatially-unresolved optical data.
  \item Another high-excitation region is located at (77\arcsec,9\arcsec)
        and has a radius of 1\farcs 5. Given the symmetry of the local 
        ionization structure, this node is probably ionized by a single star
        with a hard ionizing spectrum. The star is 0\farcs 5 towards the W of 
        the star identified as 139 by \citet{Drisetal93}. The ionization level 
        is high, with $\log(\oiiir/\hb)=0.7$. Extinction is intermediate.
\end{itemize}

        The detailed ionization structure of 
an \hii\ region is related to the issue of whether all the ionizing radiation
produced by the massive stars is processed within the nebula or whether some of 
it escapes into the more general interstellar medium of the host. 
\citet{GonDPere00} modeled the integrated spectral properties of NGC 604, 
including an analysis of both the stellar spectral energy distribution (SED)
and the photoionization of the integrated nebular spectrum. 
They concluded that, within this integrated modeling, the SED of the stars was 
adequate to account for the gas ionization and extension, and that there was no 
leak of ionizing radiation. The data presented here shows two pieces of evidence
that argue against that statement. First, the western quadrant 
of the nebula appears to be matter bounded, as seen in the ionization structure 
and diagnostic diagrams shown above. Notice that the second 
largest molecular cloud is located in this direction, which in principle might
seem incompatible with the matter bound scheme. This apparent discrepancy can
be resolved if we locate the molecular cloud ``behind'' (along the line of
sight) and the ionized gas is ``in front'', as indicated by the extinction 
analysis above, and assume that the ionized cloud is 
``broken'' so that the radiation escapes after producing the O$^{++}$ zone 
and there is no ionization front trapped in this direction. 
Second, we have measured the velocity field along the 14\arcsec-wide WHT scan 
spectrum at PA=60$\arcdeg$.  Figure~\ref{fig-vel} shows the velocity 
field of \hb\ (filled points) and of \oiiir\ (open circles), together 
with the \hb\ flux distribution (dotted line). The horizontal line marks 
the systemic velocity of $-255$ km s\m\ \citep{TenTetal00} and the horizontal
scale (in \arcsec) is centered in the main shell (seen as the two rather 
asymmetrical peaks in the flux distribution), increasing towards the SW. 
At the SW edge of the shell the gas velocity suddenly jump to 
$-270$ km s\m. This blue-shifted high-excitation ionized gas can be interpreted 
as further evidence that the shell has been broken and that the shreds have been 
blown out onto the line of sight, as suggested by \citet{TenTetal00}. 
In any case, the final word on the nature of NGC 604 as an
ionization- or matter-bounded nebula requires a complete characterization of
the young stellar population, which we will address in a subsequent paper.

\section{DISCUSSION AND CONCLUSIONS}

        A consistent picture emerges from our analysis of the gas
distribution in NGC 604: the $\approx 3.0$ Myr-old MYC
\citep{GonDPere00,Maiz00} has carved a hole in its surrounding ISM and HAS
punctured it in several places, leading to the formation of a series of
cavities and tunnels. This low-density medium is filled with hot coronal
gas that emits in X-rays and is transparent to the ionizing UV
radiation. The leftover molecular gas from the parent cloud is still
visible along several directions, but for others, including the direct
line of sight to the central part of the SOBA, it appears to have been
almost completely cleared out. What we see as a giant \hii\ region is a
composite of (a) localized high-intensity, high-excitation gas on the
surfaces of the molecular clouds directly exposed to the ionizing
radiation, and (b) a diffuse low-intensity, low-excitation component that
extends for several hundreds of pc. A similar morphology is also observed
in 30 Doradus. Both objects have a similar partition of the fluxes for
different emission lines, with most of the photons from medium and high
excitation species (e.g. \ha\ and \oiiir) being produced in the surfaces
adjacent to the molecular clouds and a more even distribution for the
photons from low-excitation species (e.g. \sii\ and \niir).  The
high-excitation regions are indeed near-bidimensional in character, given
that their thicknesses ($\sim 1$ pc) are much smaller than their extensions
(several tens of pc), so that a more appropriate name for them may be \hii\
surfaces. The morphology of the \hii\ gas in the halo is less clear: does
it fill most of the volume around the giant \hii\ region or is it
concentrated in a series of thin shells?  The complex kinematics of the
halo of both 30 Doradus \citep{ChuKenn94} and NGC 604
\citep{Yangetal96,Maiz00}, where two or more individual kinematic
components are detected in most positions, favors the second option. 
This \hii\ surface + extended halo morphology observed in NGC 604 and
30 Doradus \citep{Walbetal02a} is also consistent with what is observe in
objects at larger distances and lower spatial resolutions, such as NGC
4214-II \citep{MacKetal00}, where the low-excitation halo is easily
resolved but the \hii\ surface is reduced to a quasi-point-like high
excitation core.

        Assuming the validity of the comparison between the structure of
the giant \hii\ region in NGC 604 and in other objects such as 30 Doradus,
we can make two predictions for future observations. One is that wherever
\hii\ surfaces are present one should also detect the PDR in e.g. the NIR
H$_2$ emission lines \citep{Rubietal98b}. The second one is that if we
obtain higher-resolution images of the \hii\ surfaces, we should detect
dust pillars similar to the ones seen from the Eagle Nebula to 30 Doradus
\citep{Scowetal98,Walb02b}. Those pillars should be easier to detect where
the \hii\ surface is seen near-edge-on and should be locations where
induced star formation may be taking place. Another such place where we may
be witnessing induced star formation is the bright compact \hii\ region in
region A, which is similar to knot 1 in 30 Doradus in terms of apparent
stellar content, orientation with respect to the molecular cloud, and high
surface brightness \citep{Walbetal02a}.

        For distant giant \hii\ regions we cannot resolve the individual
components and we can only analyze their spatially-integrated properties.
Our analysis suggests that characterization of the extinction from
spatially-integrated studies might be quite uncertain, especially when the
amount of dust is relatively large.  Taking NGC 604 as an example, if we
would perform an integrated study of its properties, the underestimation of
the number of ionizing photons as derived from the $F_\alpha/F_\beta$ ratio
would be around 27\%, compared to an underestimation of around 11\% derived
from our high-spatial-resolution data. Just for comparison, if no
correction at all is applied, the underestimation increases to around
42\%. In principle, the analysis of the radio continuum could provide a
good estimate of the ionizing flux being emitted, indeed more accurate than
the value derived from the Balmer lines ratio. Nevertheless, we want to
stress that the non-thermal contribution to the radio continuum could lead
to an overestimation of the ionizing flux, unless high-resolution radio
observations are used to separate the contributions from non-thermal emitting
regions. In the case of relatively unevolved regions, like NGC 604, the
non-thermal component is negligible ($\approx 1\%$ at 3.6 cm), but it might
be significant for other objects (see \citealt{MacKetal00} for NGC 4214).
We want to stress that the uncertainties in the ionizing flux values
derived from emission lines or thermal radio continuum make it very
uncertain to derive the number of ionizing photons escaping unabsorbed from
the region, unless a careful analysis of both the stellar population and
the structure of the absorbing agents is performed.

        The reason why it is not possible to produce a simple correction
for the attenuation experienced by the gas is the intrinsically complex
geometry of a giant \hii\ region. The sources (\hii\ surfaces and halos)
and the extinction agents (the dust particles located mainly in the
adjacent molecular clouds) are extended and, to a certain degree,
intermixed. As a consequence, sources located in different points in the
plane of the sky may have quite different amounts of dust in front of
them. Furthermore, two sources along the same line of sight may have large
amounts of dust between them, as we have seen for the case of the
``flaps''. Our WFPC2 images show how extinction can rapidly vary in scales
of a pc or less and that even relatively small structures of the order of a
few tens of pc can hide a significant fraction of the
optically-unobservable \hii\ gas.  We estimate that most of the $\lesssim
11\%$ of the \ha\ flux that is not recoverable except using radio (or maybe
NIR) observations could be accounted for simply by extending the optically
bright \hii\ surfaces in the B and CD regions into an area equivalent to
the one covered by each flap, as derived from the optical geometry in
Fig.~\ref{colormosaic} (10-20 square arcseconds in each case).

This complex geometry of GHRs explains the well known effect that in many
galaxies hosting strong starbursts the average extinction experienced by
the stars is significantly lower than the integrated extinction derived
from the Balmer emission lines, as discussed e.g. by
\citet{Calzetal00}. In NGC 604 the SOBA is located in a low extinction
region because it has cleared out a cavity around it. A similar result was
obtained for NGC 4214 \citep{Maizetal98}. In the next paper we will analyze
the extinction affecting the stellar continuum of the individual stars 
in NGC 604 one by one, in more detail. 

We can conclude that in NGC 604 a large fraction of the extinction is
caused by dust associated with the GHR itself, although not evenly
distributed. A similar conclusion was reached by \citet{CaplDeha86} for
\hii\ regions in the LMC using unresolved data. It should be pointed out
that those authors indicated already almost twenty years ago that one of
the ways to continue work in this field was by using ``point-by-point
emission-line photometry''. Our analysis shows that this technique allows
indeed a better characterization of the extinction properties and, moreover,
confirms the results obtained by these authors on the LMC GHRs.

        The main cavity in NGC 604 has a diameter of $\approx 60$ pc, which
is several times smaller than the value predicted by applying the
single-star wind-blown model of \citet{Weavetal77}. According to that
model, bubble sizes should be very weakly dependent on wind luminosity or
on the density of the surrounding medium, so varying those parameters does
not provide a solution to the discrepancy. As a matter of fact, any of the
four cavities in NGC 604 could have been created by the kinetic energy
released into their surroundings by only one or a few massive stars (as
opposed to the $\approx 200$ here) in a period of 3 Myr if
\citet{Weavetal77} models were applicable here. This discrepancy is similar
to the ones found for NGC 4214 and other objects by \citet{MacKetal00} and
\citet{Maiz01d}.  We believe that the explanation lies in that the ISM is
far from being in pressure equilibrium. Recent numerical simulations by
\citet{MacL04} show that the ISM is extremely dynamic, with molecular
clouds being transient objects formed along the edges of superbubbles by
the collision of two or more of them. In those simulations, the gas
compressed in such a way can collapse to form stellar clusters but may have
only a short time to do so, since superbubbles come and go in scales of
only $\sim 10$ Myr. In other environments, such as dwarf galaxies, MYCs may
form in a different way but the requirement for a source of external
pressure should still be present. NGC 604 is known to be located at the
edge of an \ion{H}{1}-detected superbubble centered $\sim 1$ kpc towards
the SE \citep{Thil00}. According to the previously mentioned simulations,
that superbubble may have triggered the formation of NGC 604 and should
still be pushing the two largest molecular clouds associated with NGC 604
in the opposite direction to the one in which the winds and SNe from the
NGC 604 SOBA are pushing them (see Fig.~3 in \citealt{Thil00}). That source
of external pressure is likely to be one of the reasons for the
smaller-than-expected sizes of the bubbles in NGC 604.

        Two additional mechanisms that act to make superbubbles smaller
than their expected sizes at a given age may also be in effect. On the one
hand, when a giant molecular cloud is formed and once the gas becomes dense
enough, its pressure should increase locally due to self-gravity. Under
those conditions, wind bubbles from isolated stars have to fight an
additional pressure and have difficulty expanding to their expected sizes,
especially in the first stages of their development \citep{GarSFran96}.  In
a stationary isolated case, the growth of a wind-blown bubble can stall
completely. In a more realistic environment, the motion of the star should
eventually take it out of its dense birth region and its bubble may expand
to a size large enough to find an adjacent bubble. The resulting wind-wind
collision should lead to a second thermalization (the first one being
produced by the initial reverse shock wave created by the interaction of
the stellar wind with its surrounding dense medium) and, after adding more
recently-formed massive stars to the bubble, become the seed for a
superbubble. However, the whole process may be delayed by several hundreds
of thousands of years by the initial confinement of the individual bubbles,
thus contributing to the excess in the number of giant \hii\ regions
without superbubbles detected by \cite{Maiz01d}. The second mechanism that
should contribute to the small observed sizes of superbubbles lies in the
inhomogeneity of the large-scale ISM. Once a superbubble expands to a size
of $\sim 100$ pc, it is likely to encounter a density gradient in one or
several directions. The superbubble is then punctured and the hot coronal
gas flows out of the cavity through the created hole(s). The subsequent
reduction in the internal pressure of the cavity should slow down its
expansion and, eventually, halt it and even reverse it. As discussed by
\citet{TenTetal00}, the observed morphology and kinematics of NGC 604 is
consistent with such a puncture.

        In this paper we have presented our results on the spatial
distribution of the different gas phases and the dust in NGC 604. In the
next paper we will study the massive stellar population by means of HST
UV-optical photometry and HST UV objective-prism spectroscopy. The analysis
of the individual stars will provide a complete picture of the relationship
between the gas and the stars in NGC 604, and will allow to measure the
fraction of ionizing photons that potentially escape from the nebula into
the general ISM of M33.

\begin{acknowledgements}

        We would like to thank Greg Engargiola, Ed Churchwell, and Guille
Bosch for giving us access to their data. We would also like to thank
Rodolfo Xeneize Barb\'a for his help with the CO data processing and Henri
Plana for his help with the calibration of the long-slit data. This work
started as part of the GEFE collaboration. We have enjoyed the benefits
from many a discussion at different workshops and brainstorming sessions
over the years with our colleagues inside and outside of GEFE. Support for
this work was provided by NASA through grant GO-09096.01-A from the Space
Telescope Science Institute, Inc., under NASA contract NAS5-26555; by the
Spanish Government grants CICYT-ESP95-0389-C02-02, AYA-2001-3939-C03, and
AYA-2001-2089; and by the Mexican government grant CONACYT 36132-E.

\end{acknowledgements}

\bibliography{general}
\bibliographystyle{aj}

\appendix
\section*{APPENDIX: DUST MODELS}

        The ratio between the \ha\ and radio emissivities of a photoionized
region with $n({\rm He}^+)/n({\rm H}^+) = 0.09$ for case B is
\citep{CaplDeha86,ChurGoss99}:

\begin{equation}
q\frac{j_\alpha \mbox{ (erg s\m\ cm\mm)}}{j_\nu \mbox{ (Jy)}} = 
\sci{7}{90}{-10} \left(\frac{T}{10^4 \mbox{ K}}\right)^{-0.59}
\left(\frac{\nu}{10^9 \mbox{ Hz}}\right)^{0.1} ,
\label{emisratios1}
\end{equation}

\noindent where the possible errors due to the power-law approximations 
used and
the possible uncertainty in the He abundance are of the order of a few
percent.  From Eq.~\ref{emisratios1} we obtain that the (true) optical
depth at \ha\ measured from the radio and \ha\ fluxes, $F_\nu$ and
$F_\alpha$:

\begin{equation}
\taurad = \ln\left[\sci{1}{27}{9}\left(\frac{T_e}{10^4 \mbox{
K}}\right)^{0.59}
\left(\frac{\nu}{10^9 \mbox{ Hz}}\right)^{-0.1}
\frac{F_\nu (\mbox{Jy})}{F_\alpha (\mbox{erg s}^{-1}\mbox{ cm}^{-2})}\right] ,
\label{tau1}
\end{equation}

        For NGC 604, we use T = 8\,500 K
\citep{Diazetal87,GonDPere00,Esteetal02} and for the frequency of 
$\nu$ = 8.44 GHz of \citet{ChurGoss99} Eq.~\ref{tau1} becomes:

\begin{equation}
\taurad = \ln\left[\sci{9}{32}{8}
\frac{F_\nu (\mbox{Jy})}{F_\alpha (\mbox{erg s}^{-1}\mbox{ cm}^{-2})}\right] ,
\label{tau1a}
\end{equation}

        The corresponding ratio between the \ha\ and \hb\ emissivities is,
within $\sim$ 1\%:

\begin{equation}
\frac{j_\alpha }{j_\beta} = 2.859 \left(\frac{T}{10^4 \mbox{ K}}\right)^{-0.07},
\label{emisratios2}
\end{equation}

\noindent where the two emissivities are measured in the same units. From
Eq.~\ref{emisratios2} and the \cite{Cardetal89} law for $R_V = 3.2$ we obtain 
the optical depth at \ha\ measured from the ratio of the two Balmer fluxes, 
$F_\alpha$ and $F_\beta$:

\begin{equation}
\taubal = 2.42 \ln\left[\frac{F_\alpha/F_\beta}{2.859}
\left(\frac{T_e}{10^4 \mbox{ K}}\right)^{0.07}\right] ,
\label{tau2}
\end{equation}

\noindent which for T = 8\,500 K becomes:

\begin{equation}
\taubal = 2.42 \ln\left[\frac{F_\alpha/F_\beta}{2.892}\right] .
\label{tau2a}
\end{equation}

        One of the largest uncertainties in measuring \taubal\ comes
from the assumed value of $R_V$. For example, values of 2.6 or 4.4 change the 
constant that multiplies the logarithm in Eq.~\ref{tau2} to 2.04 and 3.07, 
respectively. Here we will ignore this question but we plan to study it in
future works.

        The standard attenuation model is that of a uniform foreground
screen model. There we have that \taurad=\taubal, since all areas observed
are affected by dust in the same degree and no scattering is injecting
photons back into the line of sight.

        A uniform foreground screen is not very realistic for \hii\
regions, since images reveal the existence of localized dust clouds. Here
we explore a model in which we assume that the dust in the aperture is
distributed in a patchy foreground screen, in such a way that the area
covered is $\gamma$ times the area free of dust and the screen yields an
optical depth at \ha\ of \taucov\ for the areas behind it. In this case,
the values of \taurad\ and \taubal\ for the aperture will be weighted means
of \taucov\ (the optical depth experienced by the areas covered by dust)
and zero (the optical depth experienced by the areas free of dust).

\begin{figure}
\centerline{\includegraphics*[width=\linewidth]{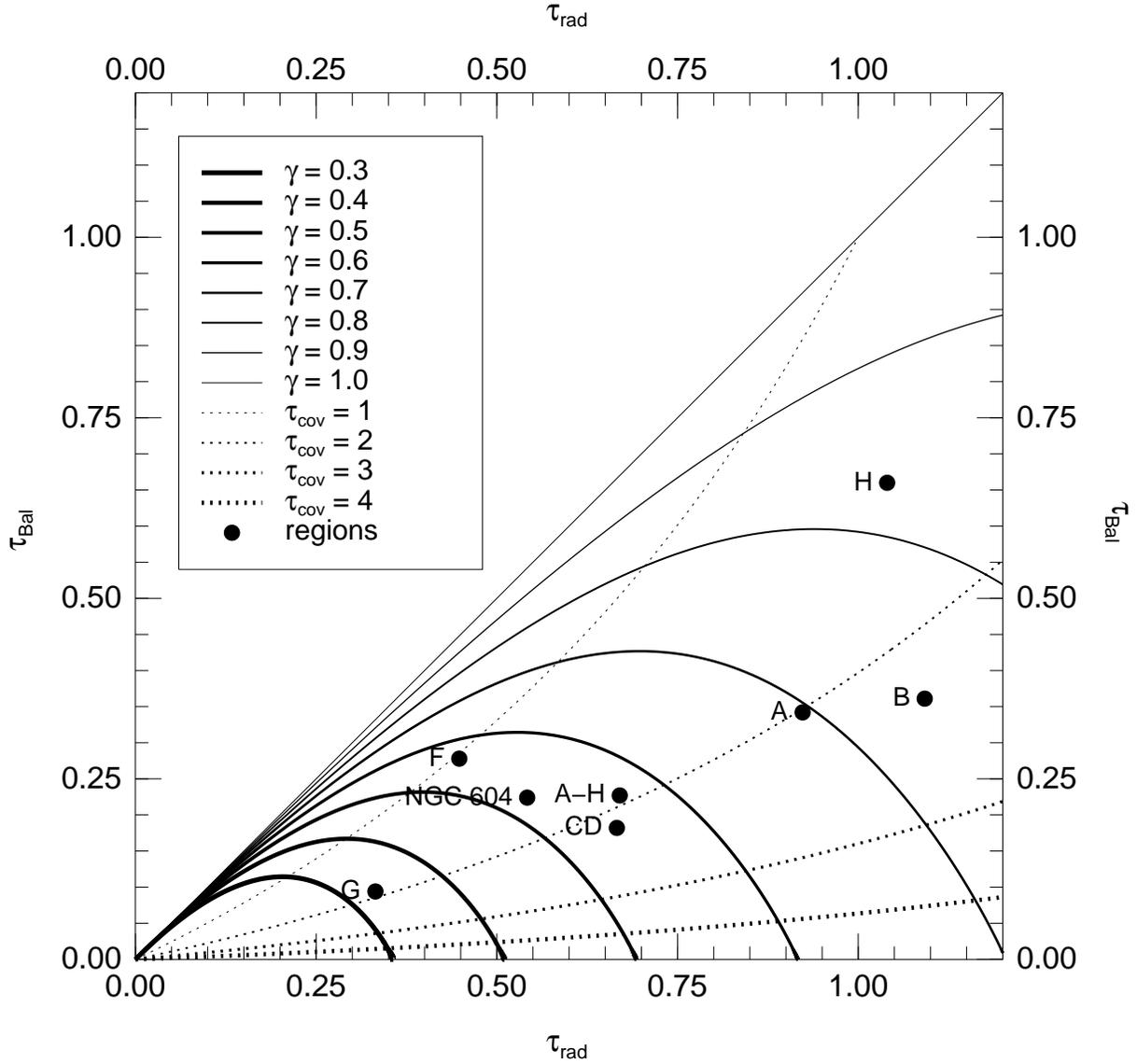}}
\caption{\taubal\ vs. \taurad\ for a patchy foreground model with different
values of $\gamma$, the area covered by the screen. Dashed lines show the
location of different values of the optical depth of the screen. The values 
obtained for our regions are also plotted.}
\label{extplot}
\end{figure}

\begin{figure}
\centerline{\includegraphics*[width=\linewidth]{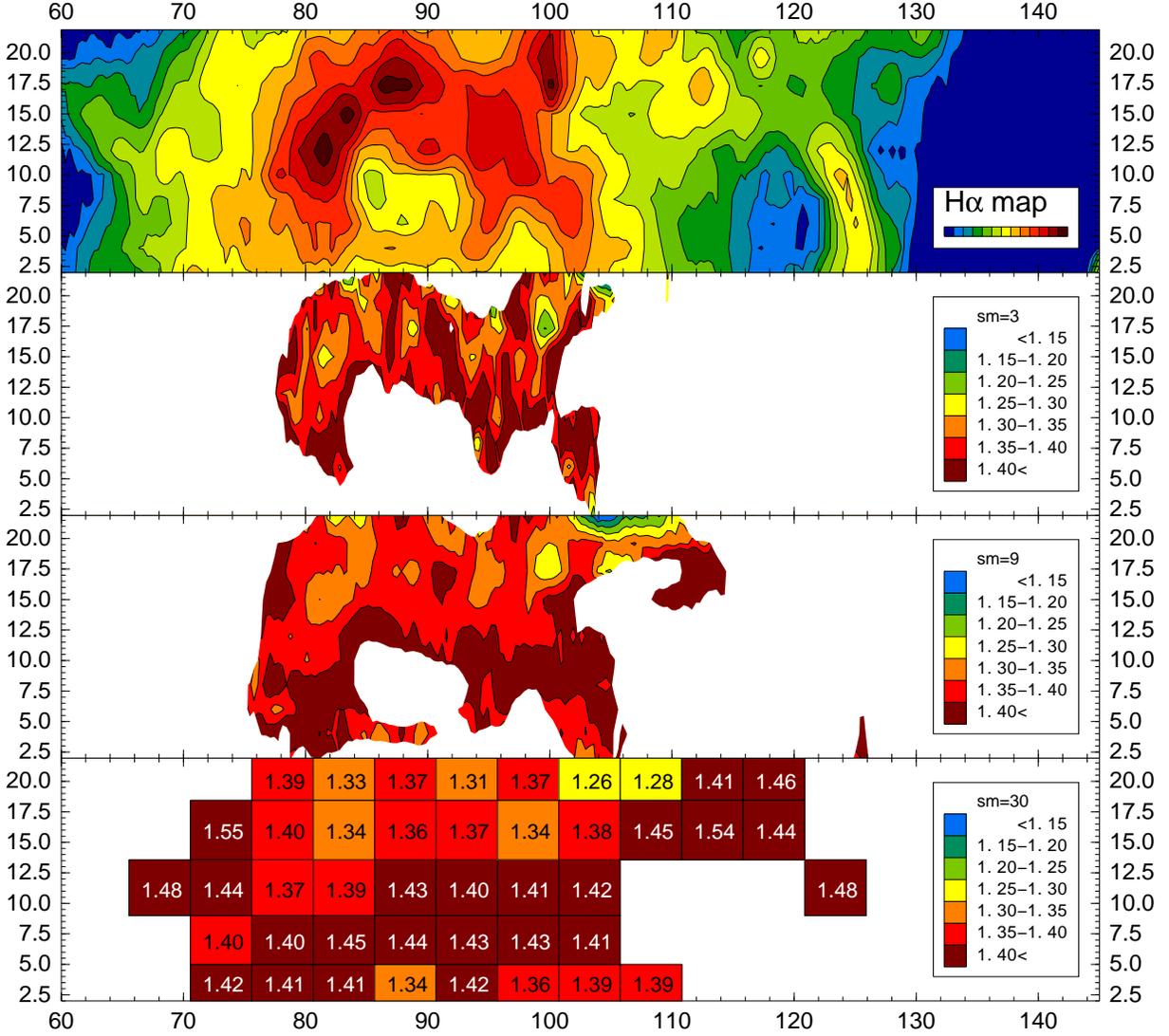}}
\caption{Synthetic \ha\ map (top panel), density ratio maps (two middle panels),
and density ratio table (bottom panel) produced from the long-slit data. The 
first density ratio map has been smoothed over 3 pixels (1\arcsec) along each 
slit while the second density ratio map has been smoothed over 9 pixels 
(3\arcsec). The color-coded density ratio table shows data averaged over 2 
adjacent long slits and 15 pixels along the slit direction. In each of the 
lower three panels, data is shown where the uncertainty of the ratio is less 
than 0.08 in order to ensure that the values shown are relevant. The coordinates
are expressed in arcseconds with north towards the bottom and east towards the
right (see Fig.~\ref{longslits}) with the first slit centered at $y=1\arcsec$ 
and its first pixel at $x=1/3\arcsec$. In these coordinates 
$(x,y) = (93\farcs4,14\farcs25)$ corresponds to 
($1^{\rm h}\,34^{\rm m}\,33^{\rm s},\; 30\arcdeg\,47\arcmin)$ in J2000.}
\label{slosrmaps}     
\end{figure}

\begin{figure}
\centerline{\includegraphics*[width=0.49\linewidth]{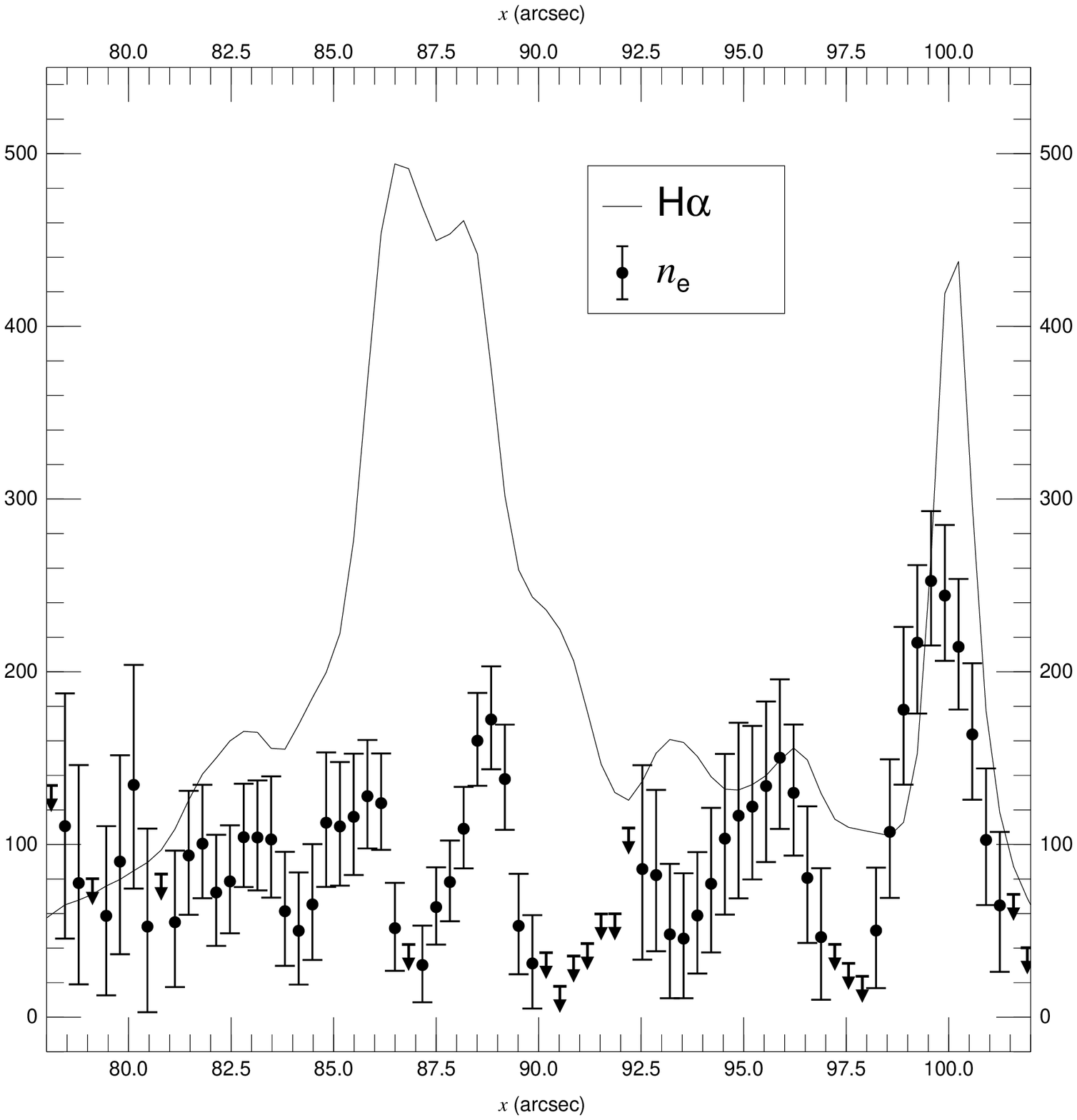} \
            \includegraphics*[width=0.49\linewidth]{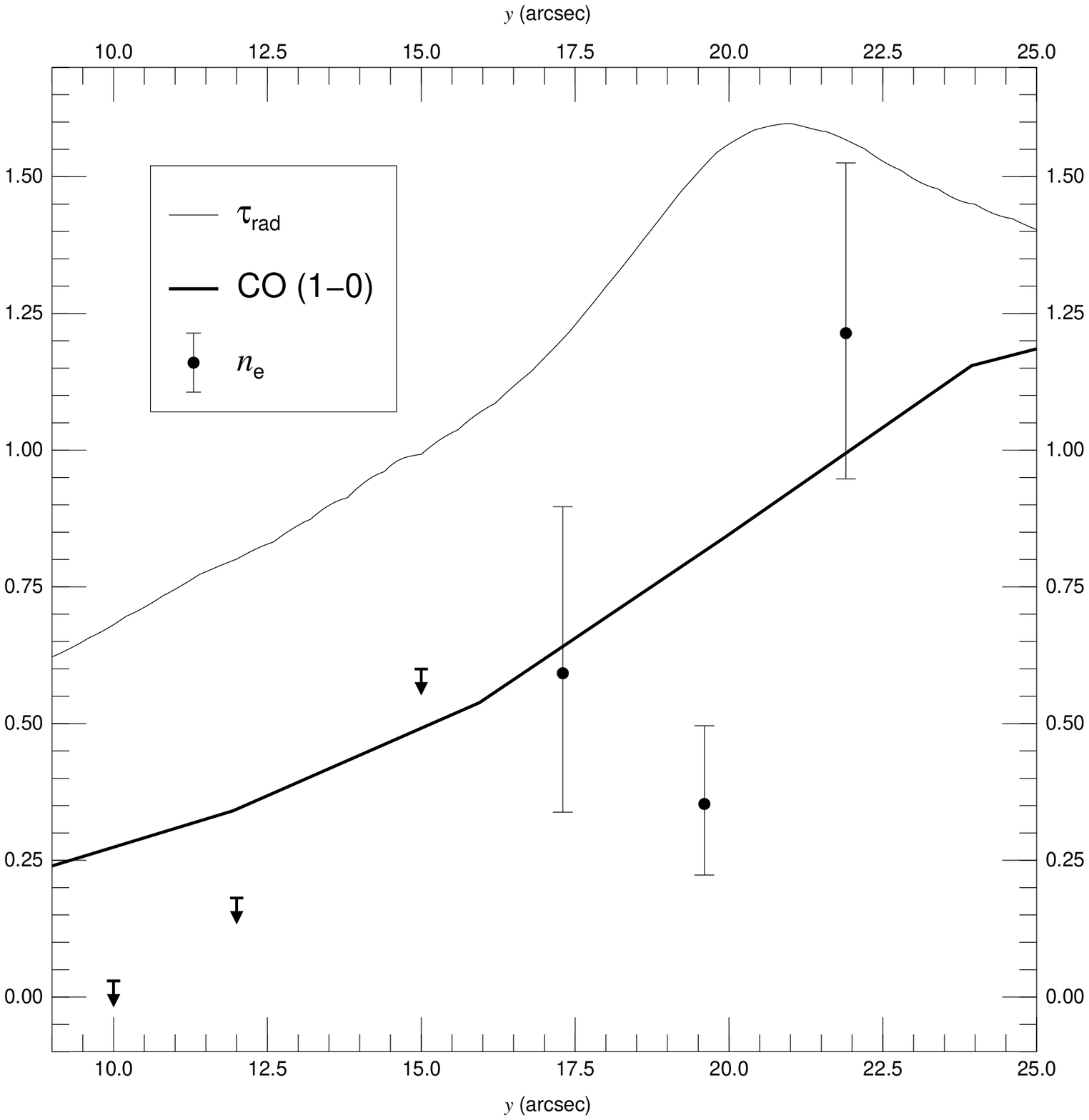}}
\caption{(left) H$\alpha$ flux and electron density at $y=17\farcs 3$
derived from slit data as a function of the horizontal coordinate established in
Fig.~\ref{slosrmaps}. The density has been smoothed with a 3-pixel 
(1\arcsec) box. H$\alpha$ is expressed in units of 
$20\cdot 10^{-17}$ erg s$^{-1}$ cm$^{-2}$ arcsec$^{-2}$ and $n_e$ is expressed
in cm$^{-3}$. (right) True optical depth at H$\alpha$, CO ($1\rightarrow 0$) 
intensity, and 
electron density at $x=105\farcs 3$ as a function of the vertical coordinate 
established in Fig.~\ref{slosrmaps}. \taurad\ is obtained from 
\citet{ChurGoss99}, the CO data is from \citet{Engaetal03}, and $n_e$ is 
derived from slit data. The density has been smoothed with a 9-pixel 
(3\arcsec) box. The 
value of \taurad\ can be read directly from the labels in the vertical axis,
the CO scale is arbitrary, and $n_e$ is expressed in units of 300 cm$^{-3}$.}
\label{cuts1}     
\end{figure}

\begin{figure}
\centerline{\includegraphics*[width=0.49\linewidth]{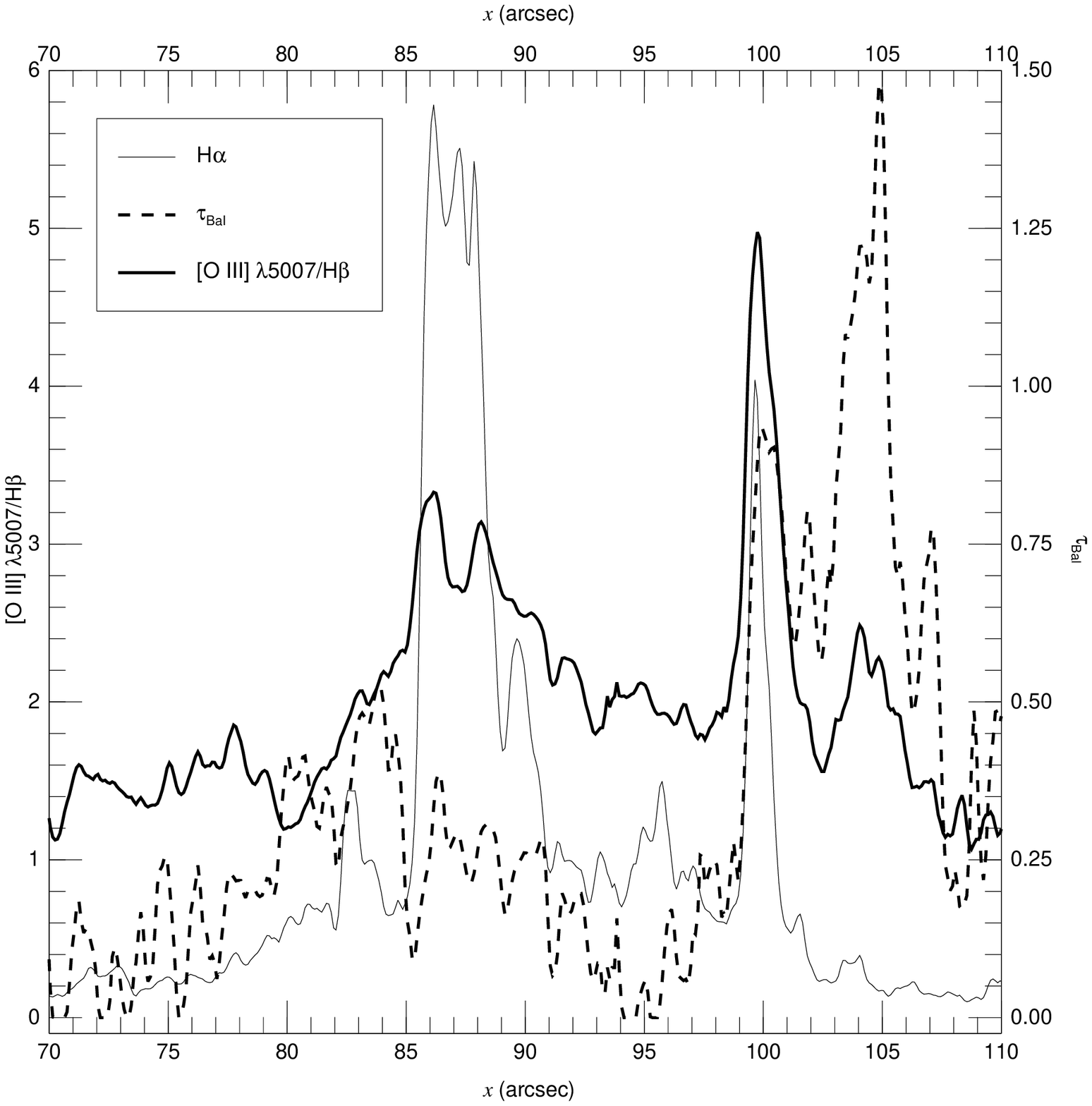} \
            \includegraphics*[width=0.49\linewidth]{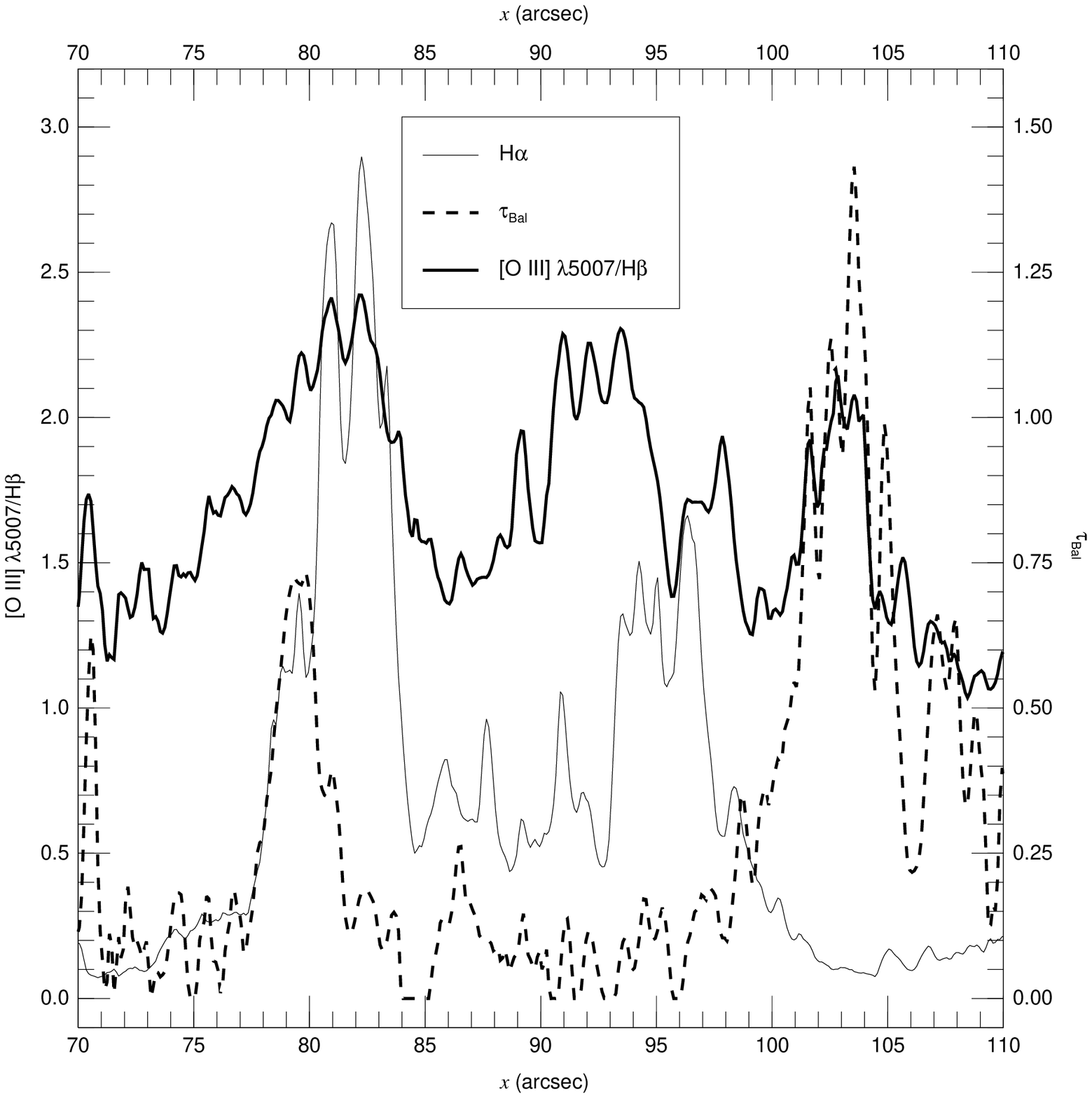}}
\caption{H$\alpha$ flux, \taubal, and \oiiir/\hb\ measured from WFPC2 data
as a function of the horizontal coordinate established in Fig.~\ref{slosrmaps}.
The left figure corresponds to $y=16\farcs 6$ and the right one to
$y=13\farcs 4$. H$\alpha$ is shown in arbitrary units while the scale for 
\taubal\ and \oiiir/\hb\ can be read from the left and right sides,
respectively.}
\label{cuts2}     
\end{figure}

\begin{figure}
\centerline{\includegraphics*[angle=90,width=\linewidth]{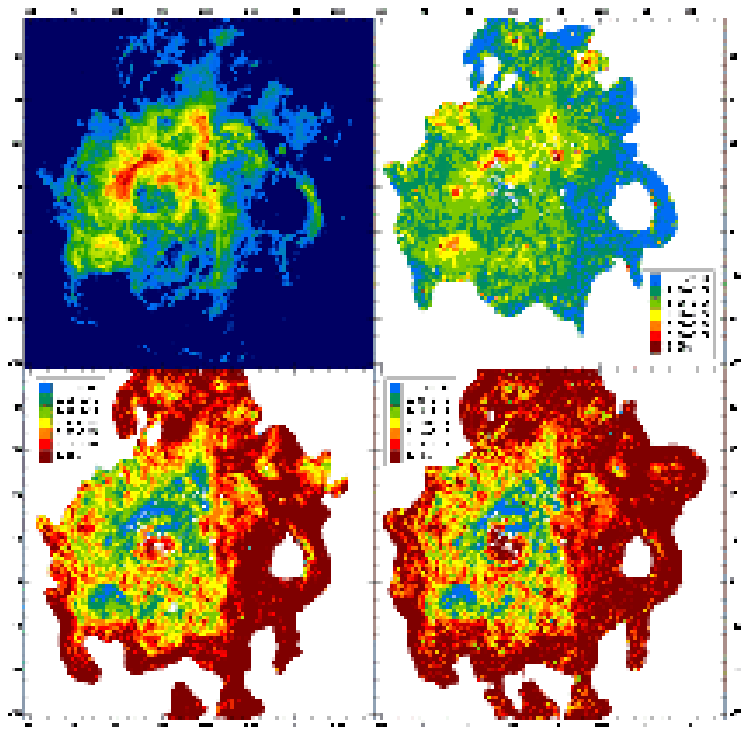}}
\caption{A continuum-subtracted \ha\ WFPC2 image of NGC 604 is shown in the
upper left panel.  The other three panels show excitation maps of the same
region: \oiiir/\hb\ (upper right) \sii/\ha\ (lower left), and \niir/\ha\
(lower right). Areas in white have been masked due to strong stellar 
contamination or low signal-to-noise.  The field is the same as the one in 
Fig.~\ref{wfpc2siioiiir}.  See Fig.~\ref{slosrmaps} for an explanation of the 
coordinate system.}
\label{wfpc2ratios}
\end{figure}

\begin{figure}
\centerline{\includegraphics*[angle=270,width=\linewidth]{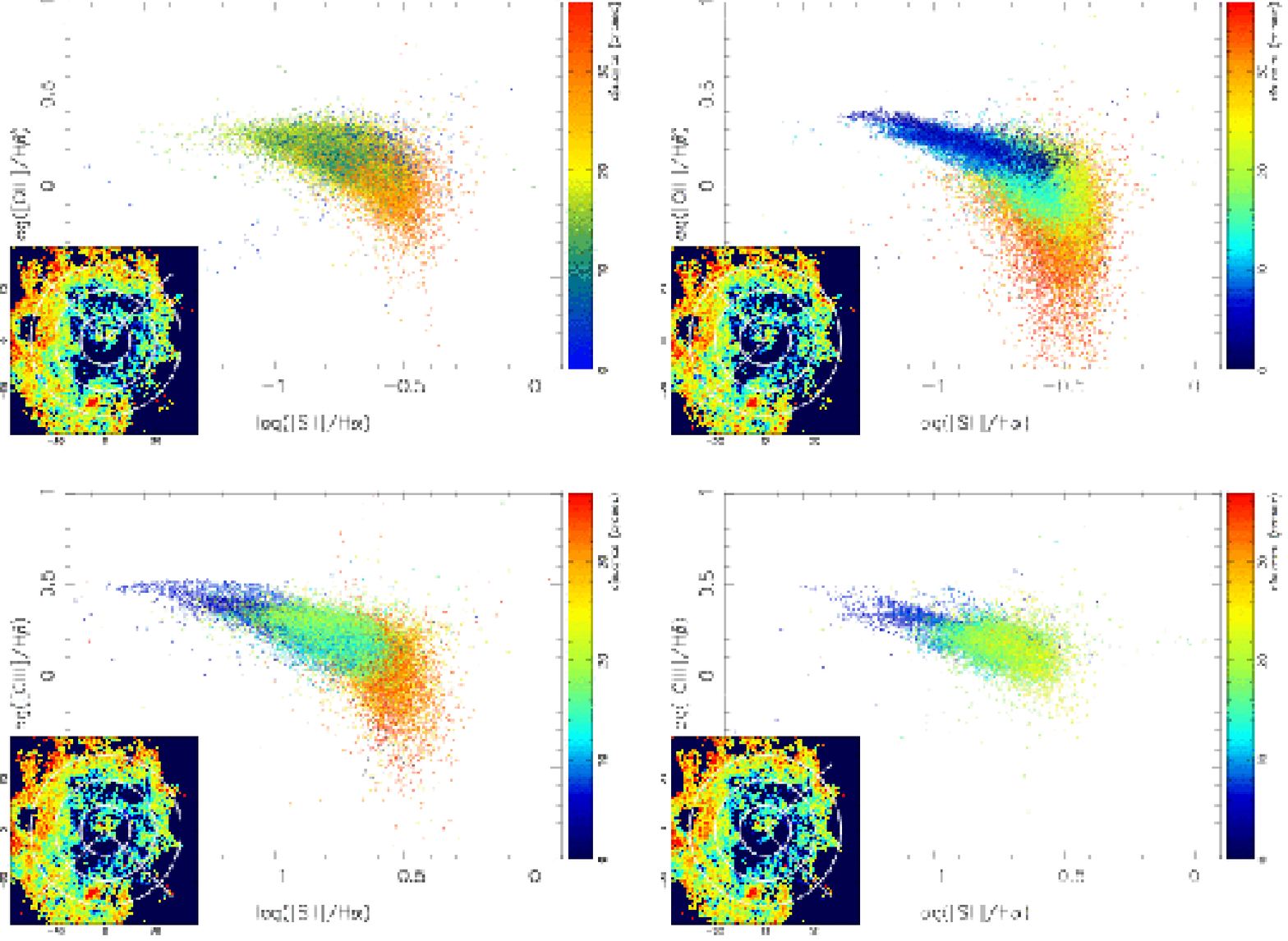}}
\caption{CIS pixel-by-pixel diagnostic diagram, log(\oiiir/\hb) vs. 
log(\sii\ha), for the four quadrants ($\pm 45\arcdeg$ centered on the four 
cardinal points, from left to right and from top to bottom, N, E, S, and W). 
The inset in each panel shows the \sii/\ha\ image with 
superposed circles at radii of 10\arcsec, 20\arcsec, and 30\arcsec, and two 
lines indicating the quadrant plotted. The overall loci of the points in the 
diagram mark the general features of the CIS, and it is apparent that while 
the three diagrams corresponding to N$\pm 45\arcdeg$, E$\pm 45\arcdeg$,
and S$\pm 45\arcdeg$ reach to very low values of log(\oiiir/\hb) $<-0.5$ at 
large distances, the W$\pm 45\arcdeg$ diagram has all its pixels in the high 
to intermediate excitation regime, log(\oiiir/\hb) $>-0.2$. This is strong 
evidence that the nebula is matter bounded towards the West.}
\label{diagn}
\end{figure}

\begin{figure}
\centerline{\includegraphics*[width=0.58\linewidth]{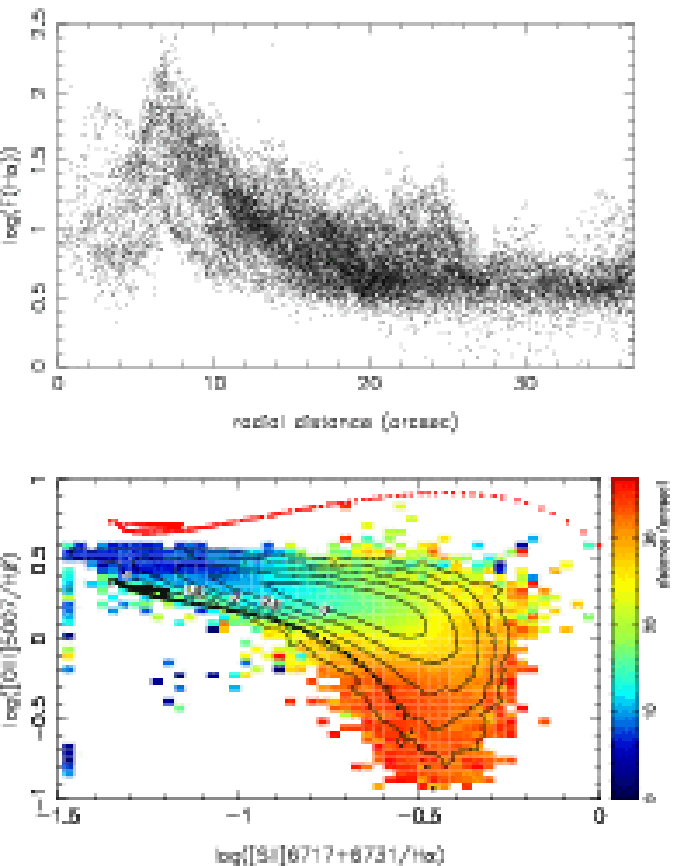}}
\caption{The top panel shows the radial distribution of the CIS \ha\ flux (in 
log scale). The general appearance is that of an empty shell integrated along 
the line of sight, with an inner radius of 8\arcsec, given by the peak flux, 
plus an extended low surface brightness halo. 
The lower panel shows the log(\oiiir/\hb) vs. log(\sii/\ha) 
diagnostic diagram of the CIS. This diagram has been constructed
from the pixel-by-pixel diagnostic diagram of the points belonging to the CIS,
excluding those regions belonging to SISs.
Each square represents the density of individual points located
in that part of the diagram. The contours give the density of points in log 
scale (i.e., the contour labeled 3 indicates that that part of the diagram is
populated by 1000 individual image pixels). The color codes for the average 
radial distance of those points to the CIS center. Two photoionization models 
(calculated using Cloudy) similar to those described in the 
simple approach followed by \citet{GonDPere00} are plotted,
the black points for an age of 2.75 Myr, and the red ones for an age of 3 Myr.
The density distribution used is taken as the azimuthal average rms density 
distribution as obtained from the \ha\ flux, and converted to actual density 
via the filling factor $\phi$ (taken as a parameter, and equal to 0.1 in the 
plot).}
\label{CIS-shell}
\end{figure}

\begin{figure}
\centerline{\includegraphics*[angle=270,width=\linewidth]{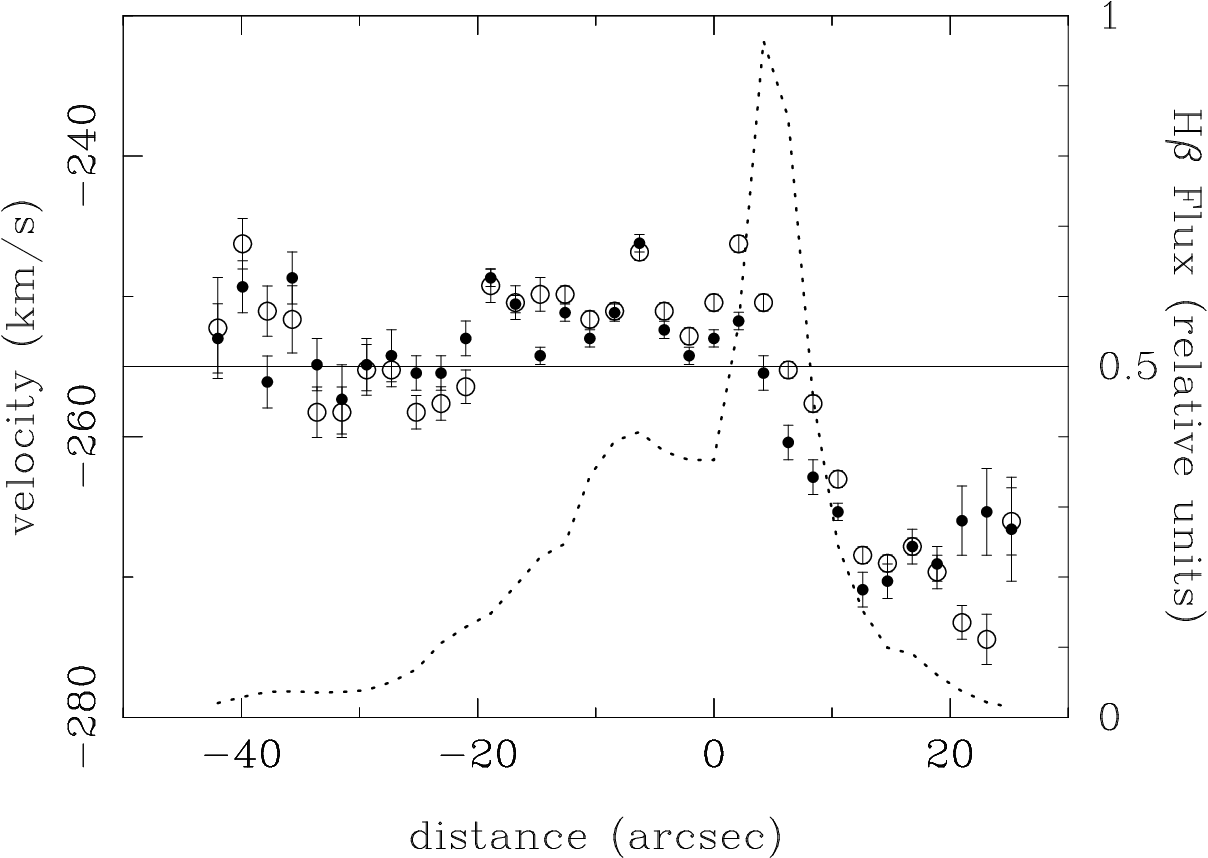}}
\caption{Velocity field of \hb\ (filled points) and of \oiiir\ (open circles), 
together with the \hb\ flux distribution (dotted line). The horizontal line 
marks the systemic velocity of $-255$ km s\m\ \citep{TenTetal00}. The 
abcissa scale is centered in the main shell (seen as the two rather 
asymmetrical peaks in the flux distribution) and increases towards the SW. 
At the SW edge of the shell the gas velocity suddenly jump to $-270$ km s\m. 
This blue-shifted high-excitation ionized gas can be interpreted as further 
evidence that the shell has been broken and that the shreds have been blown out 
onto the line of sight, as suggested by \citet{TenTetal00}.}
\label{fig-vel}
\end{figure}

        Our results are shown in Fig.~\ref{extplot}, where we represent
\taubal\ as a function of \taurad\ for different values of $\gamma$.  As
expected, \taubal\ is always smaller than \taurad\ when the dust is not in
a uniform foreground screen and scattering is not relevant (see
e.g. \citealt{CaplDeha86}. For our patchy foreground screen model, \taubal\
first increases while the obscured areas contribute significantly to the
detected radiation. When the amount of dust present in the screen becomes
large enough, the curve acquires a negative slope and ends up going back to
zero since the dominant effect is to let only the unobscured regions
contribute to the detected Balmer photons. Note that the region defined by
$\taubal \ge 0$, $\taurad \ge \taubal$ is completely covered by the family
of models for $\taucov \ge 0$, $0\le \gamma \le 1$, with a one-to-one
mapping which is only degenerate for $\taubal=\taurad=0$.

\end{document}